Fig. 1 σ²/N vs. z, without evolution

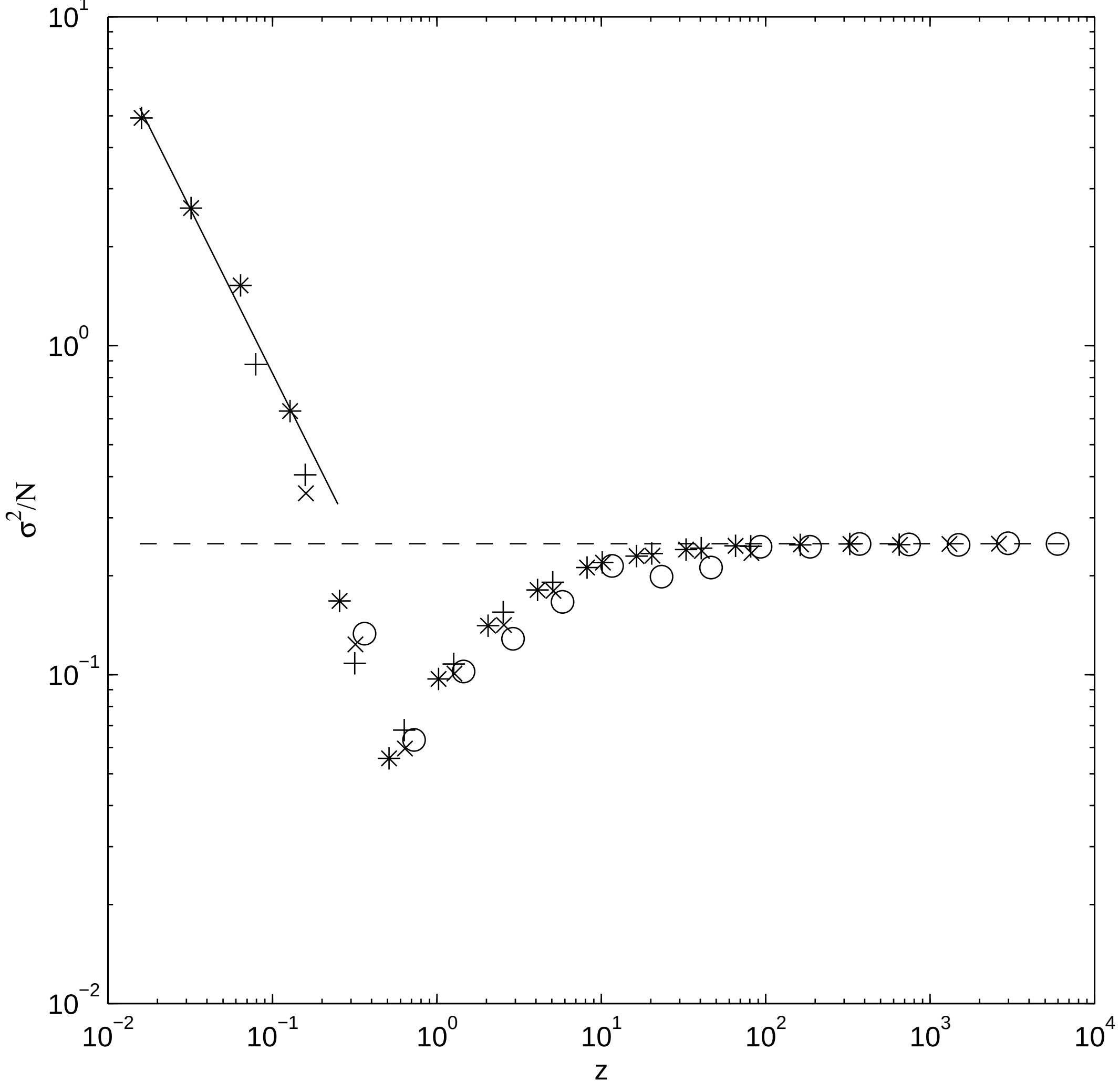

Fig. 2 $\sigma^2/N$ vs. m for p=20%

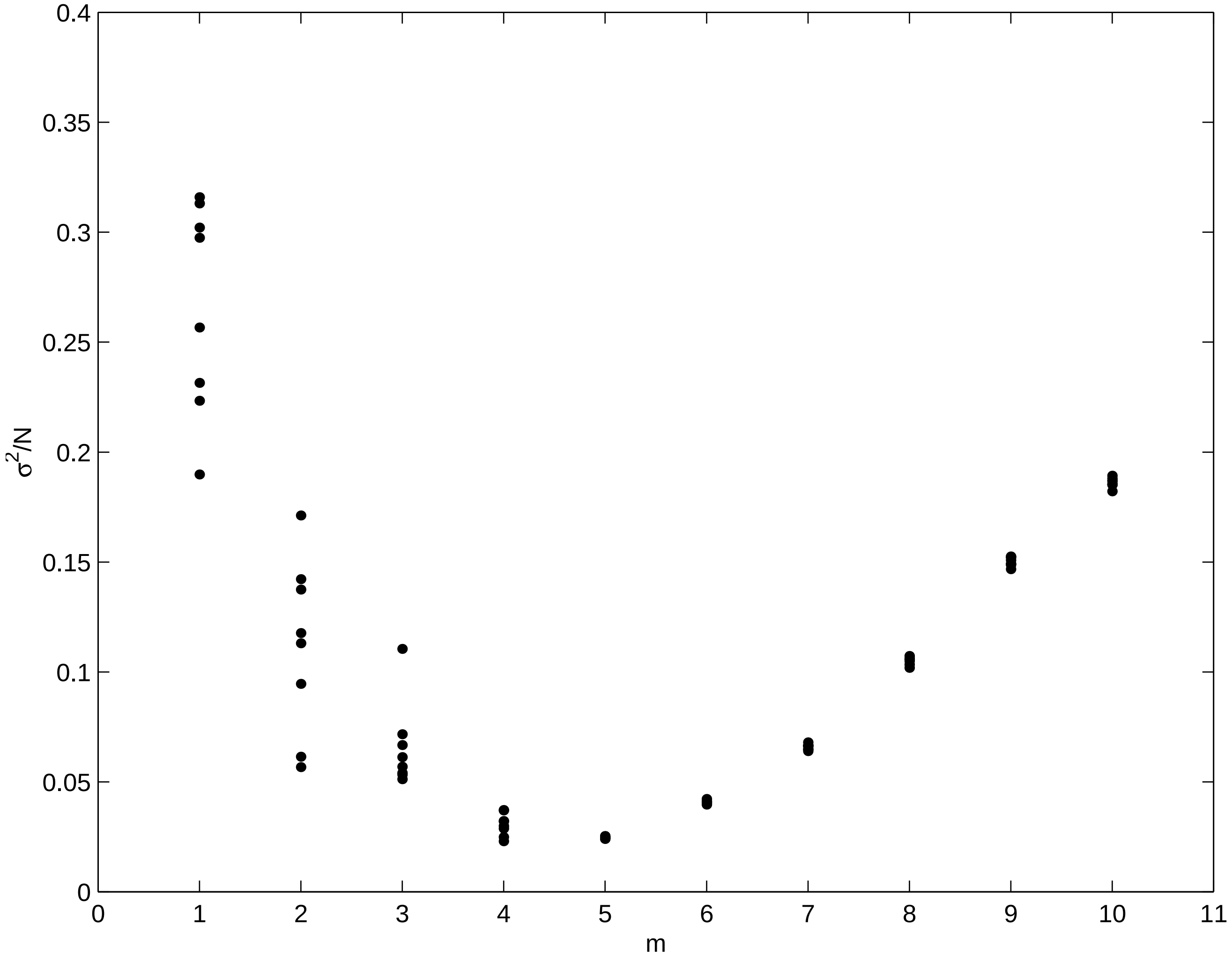

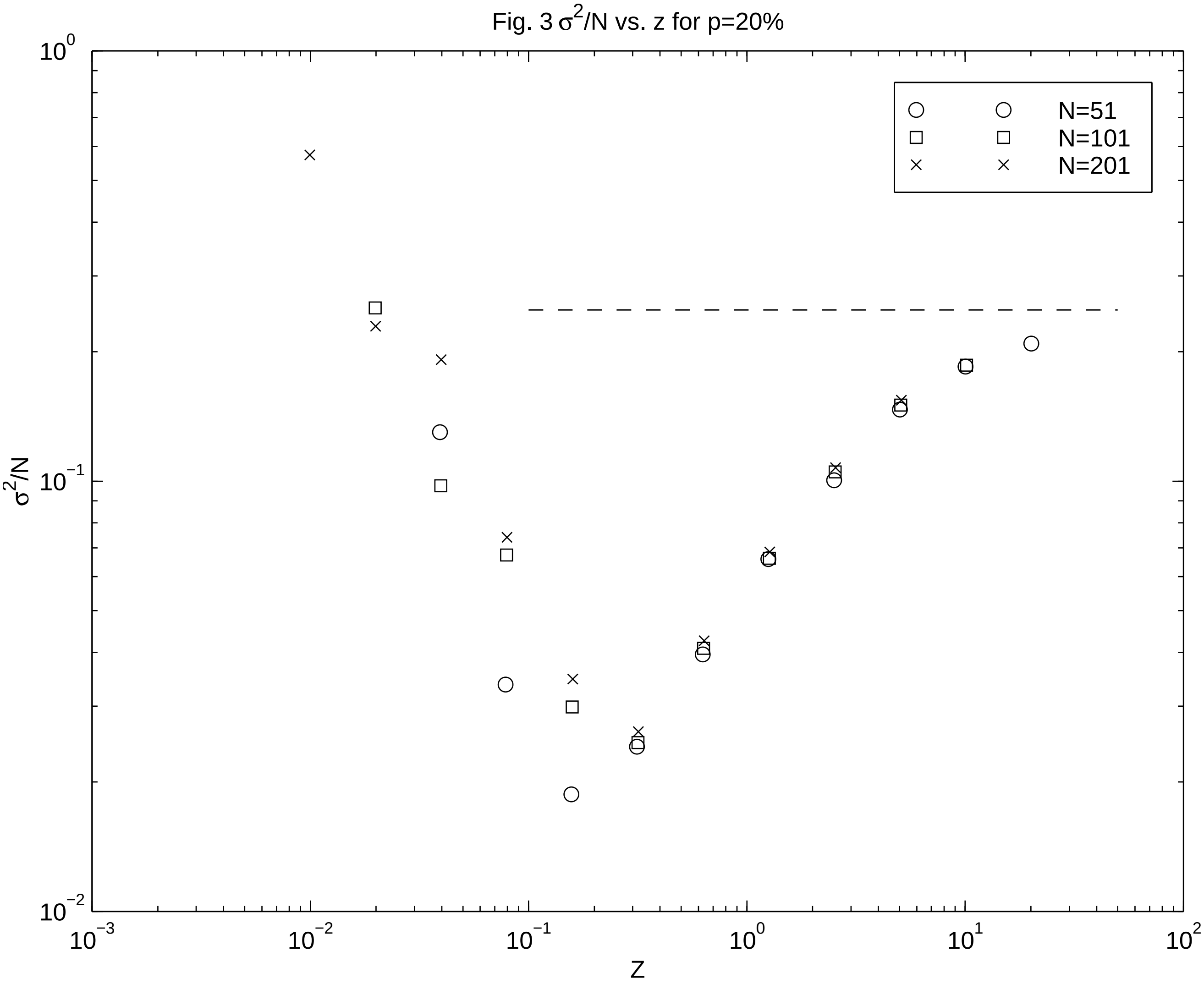

Fig. 4 $\sigma^2/N$ vs. m for different poverty levels, N=101

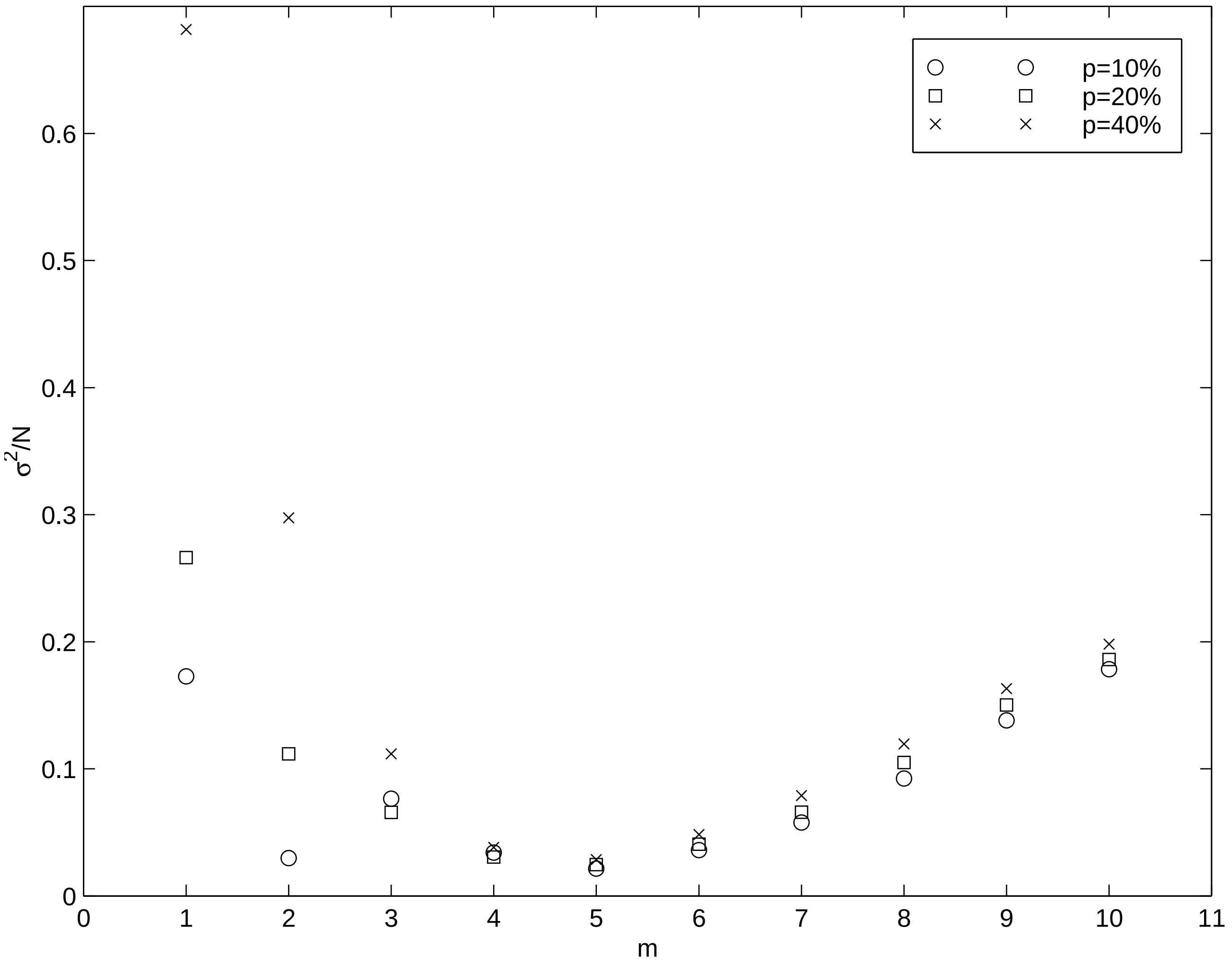

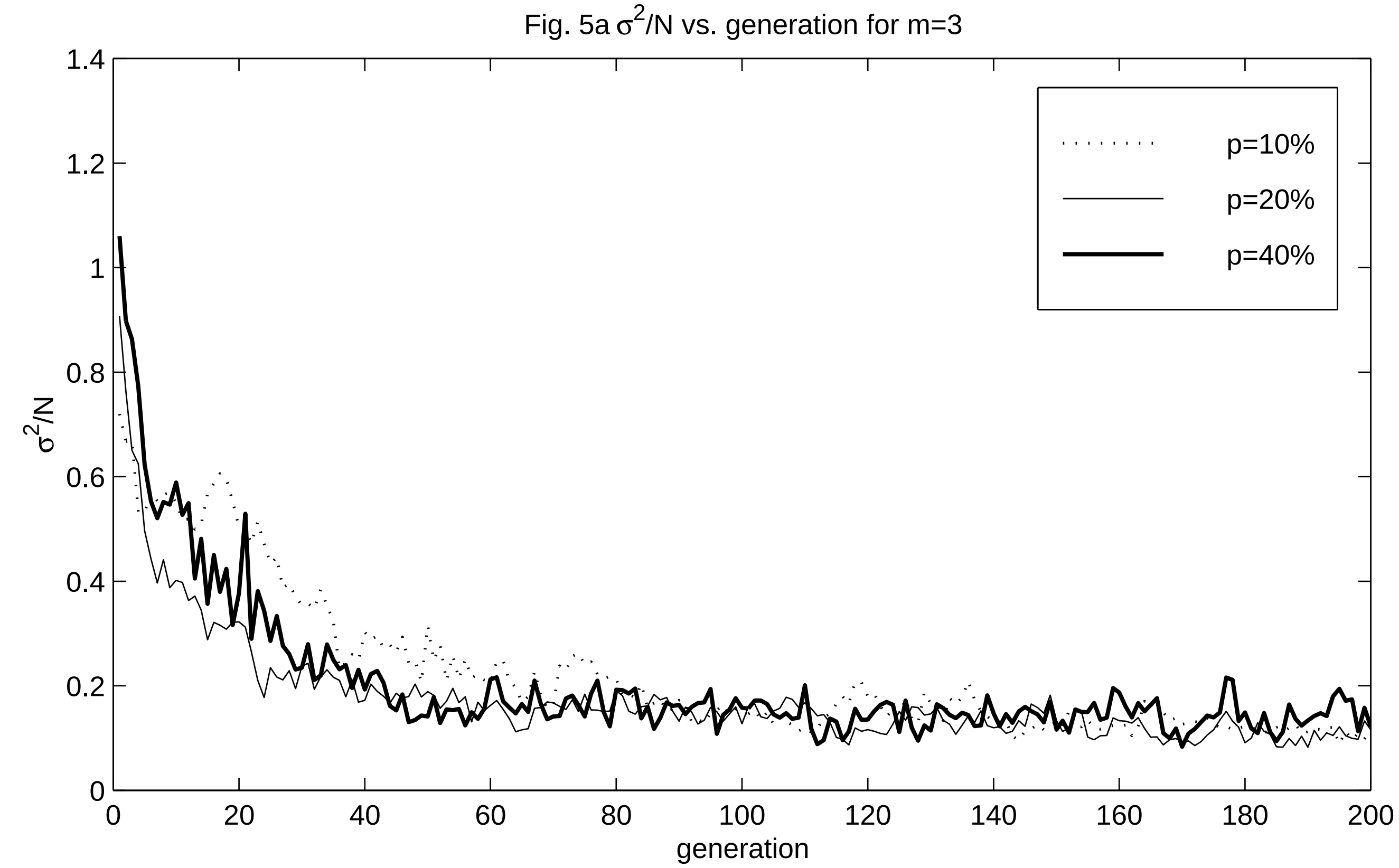

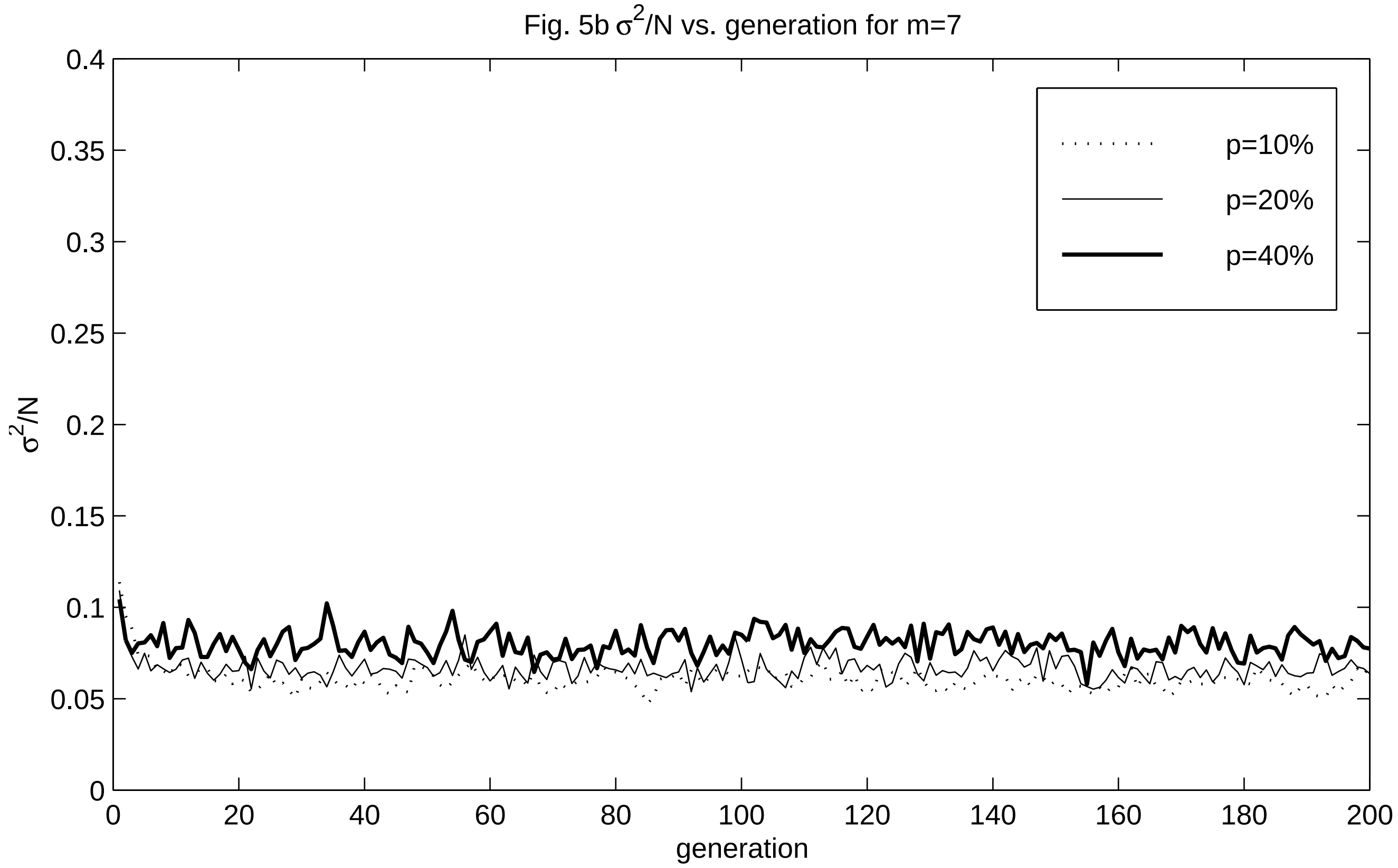

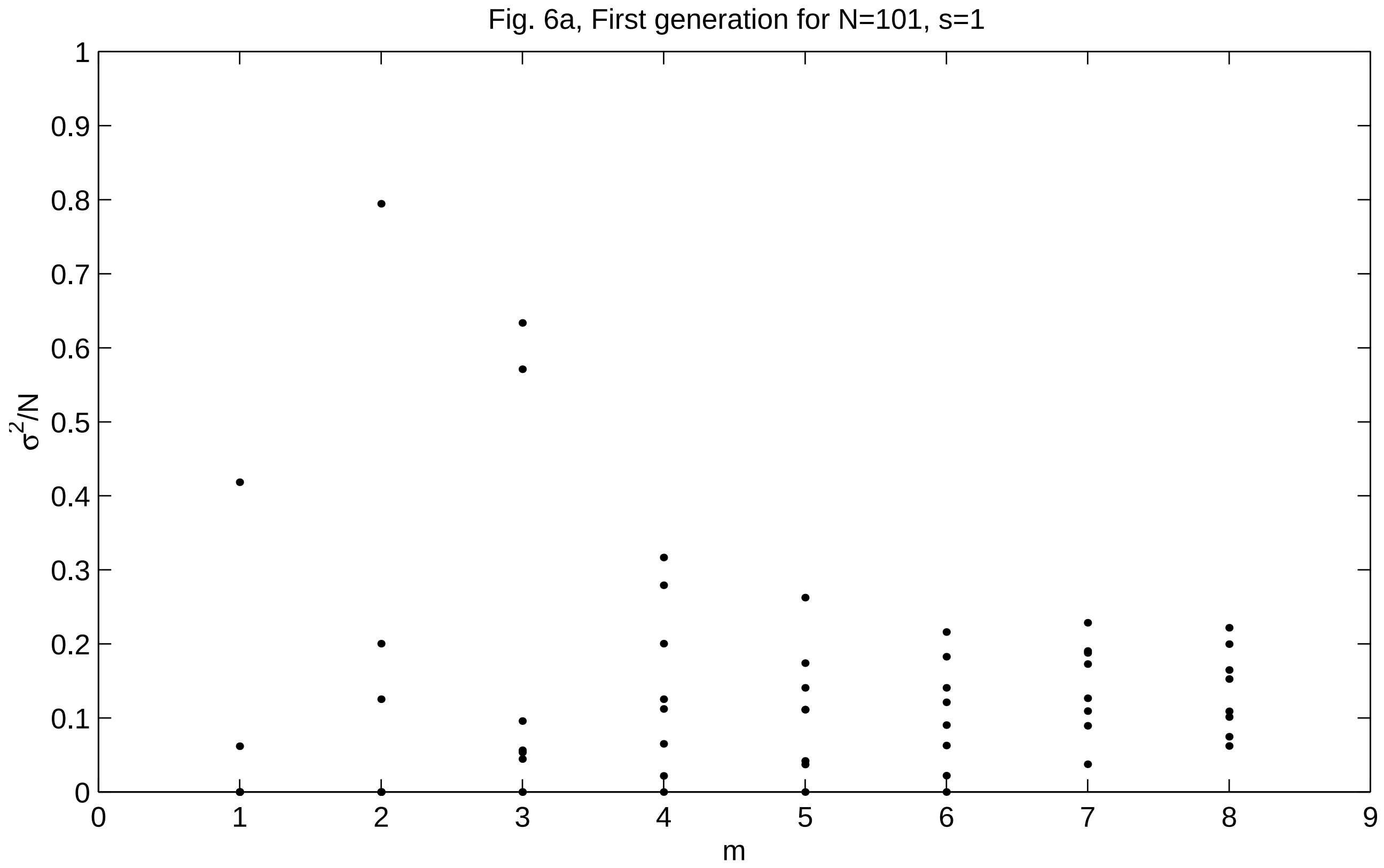

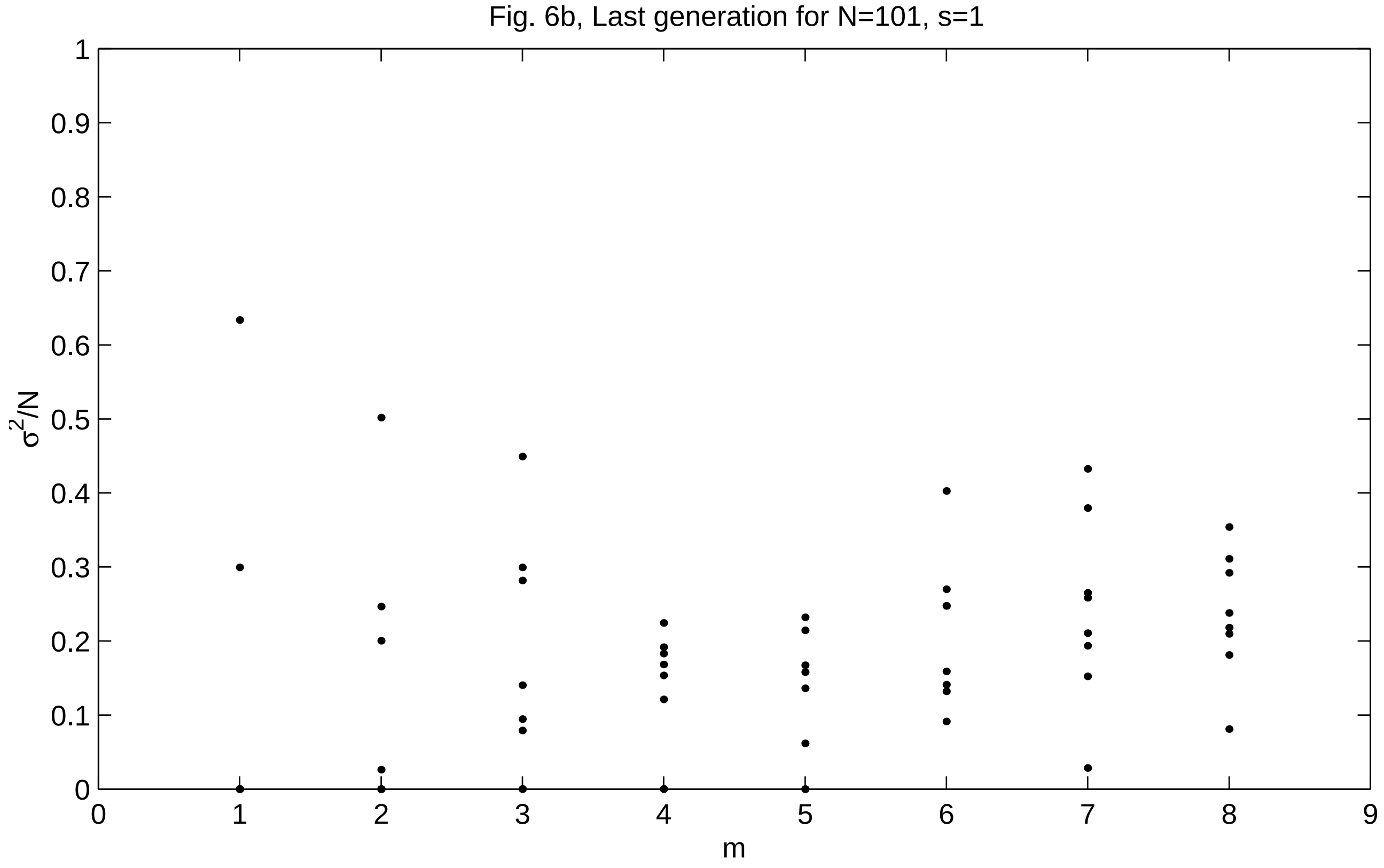

Fig. 7a $<D_h>$ vs. generation for m=3

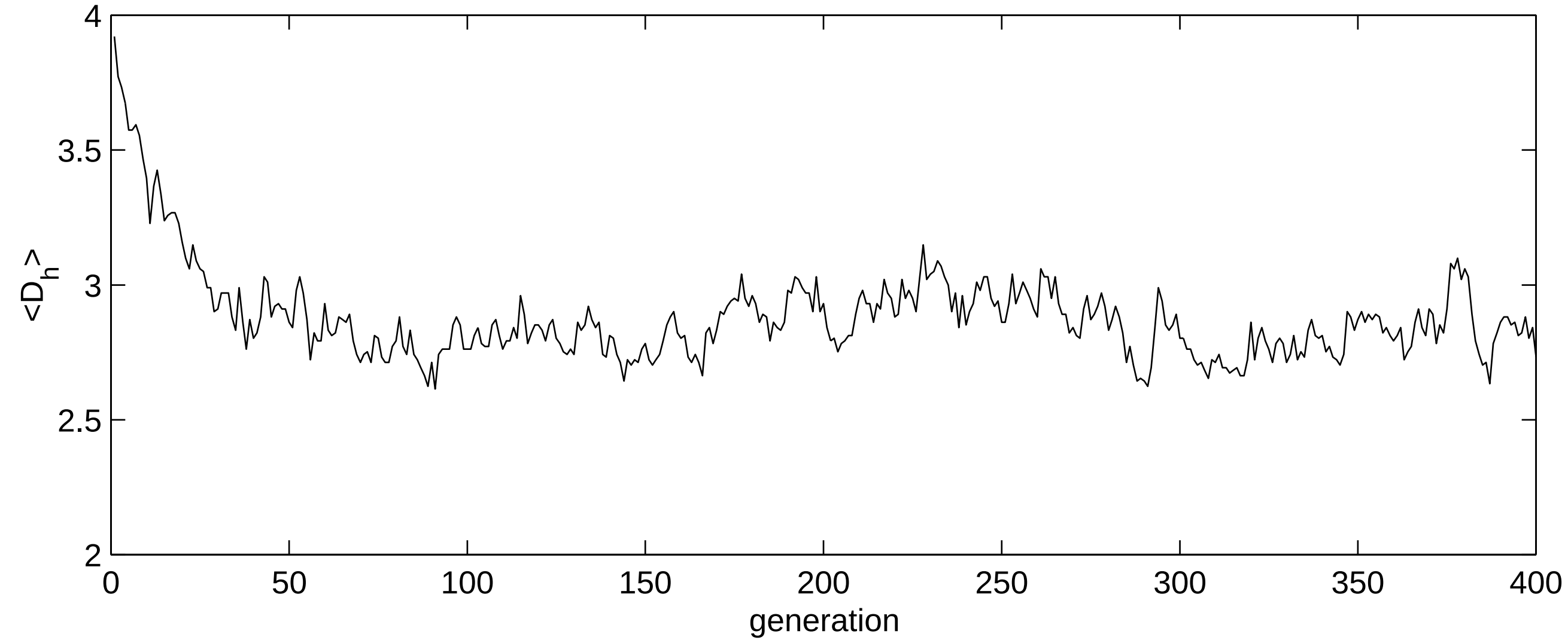

Fig. 7b $<D_b>$ vs. generation for m=3

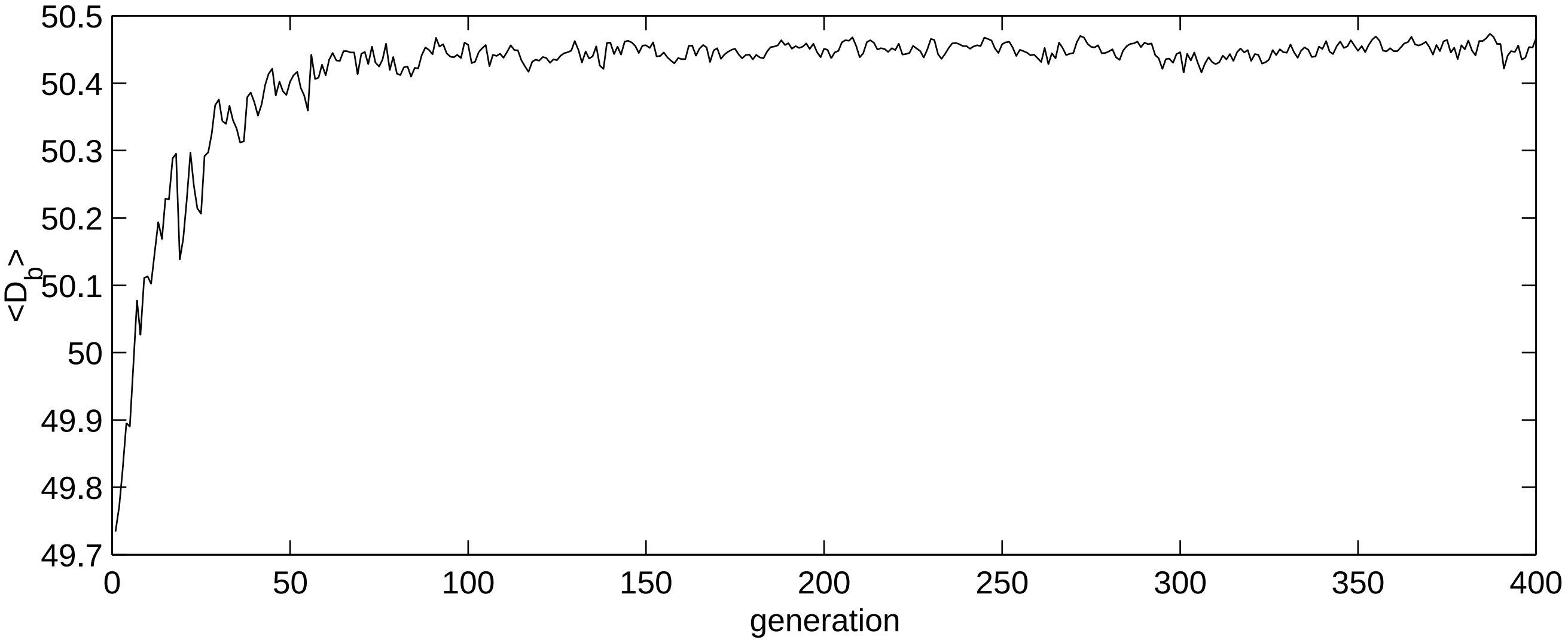

Fig. 7c $\sigma^2/N$ vs. generation for m=3

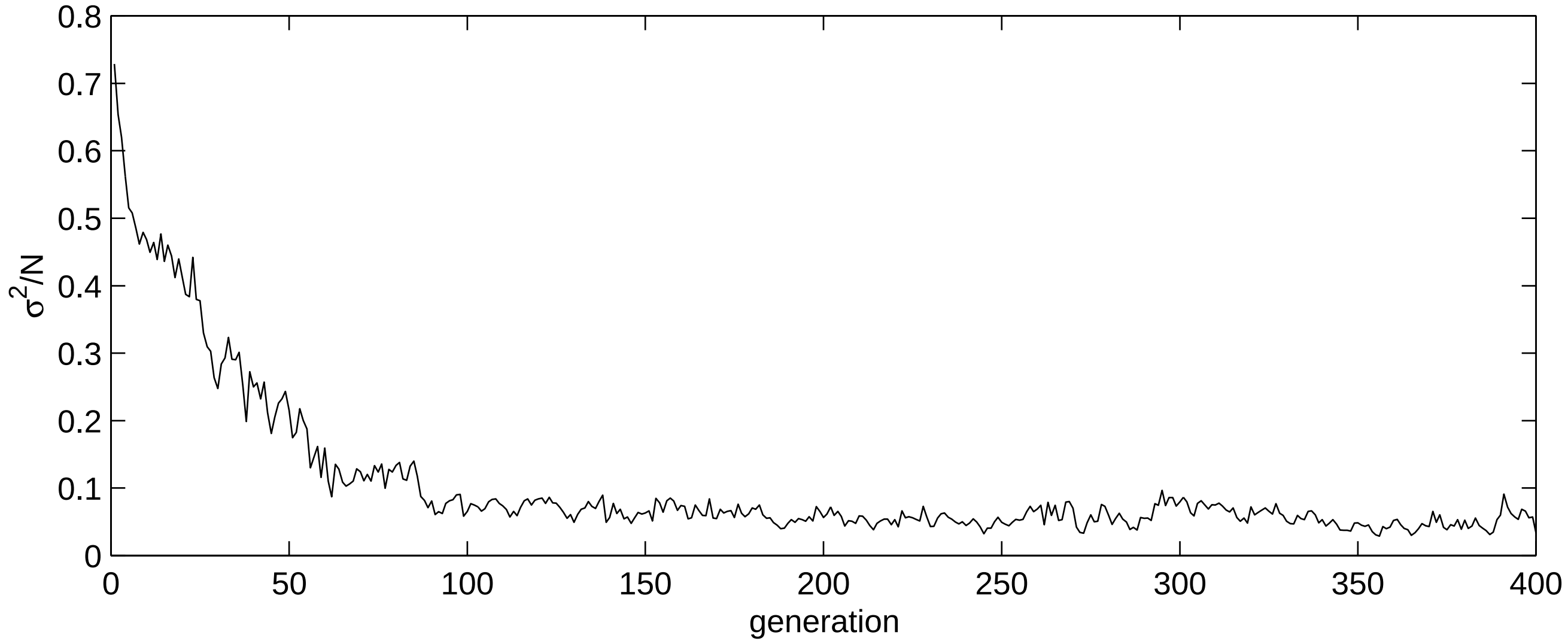

Fig. 8a $P(1|u_m)$ for the 1st generation for m=3, N=101

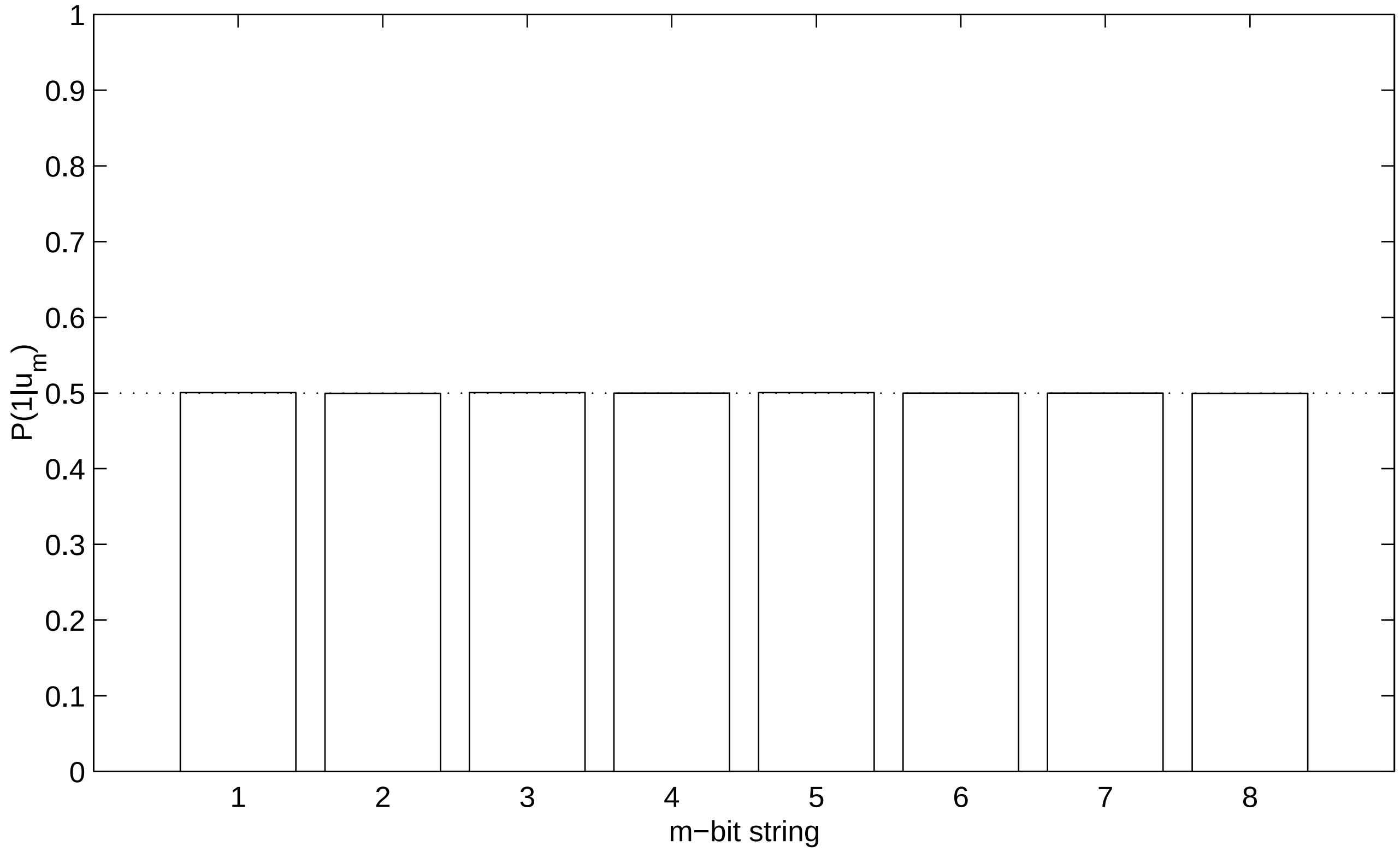

Fig. 8b $P(1|u_m)$ for the 400th generation for m=3, N=101

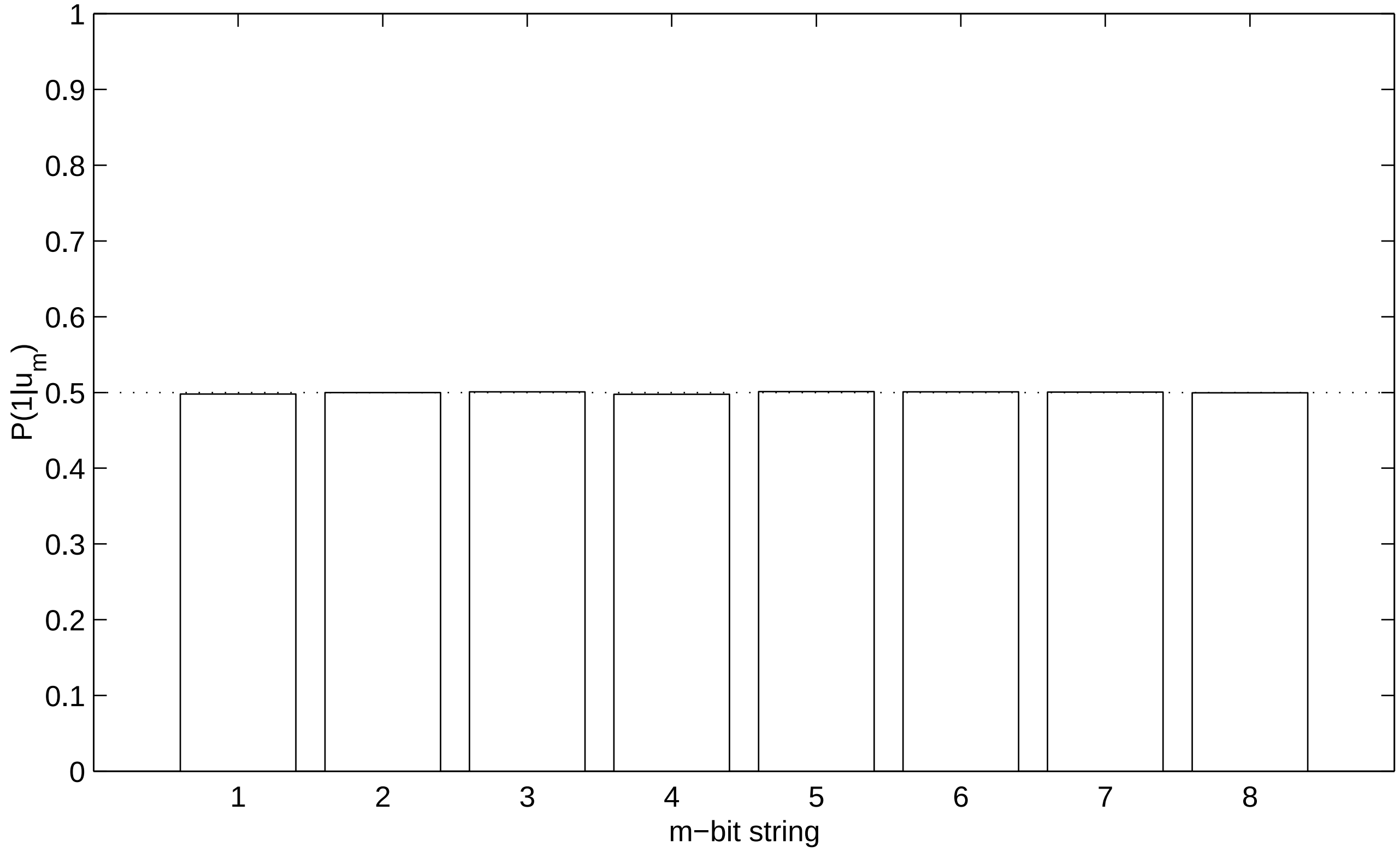

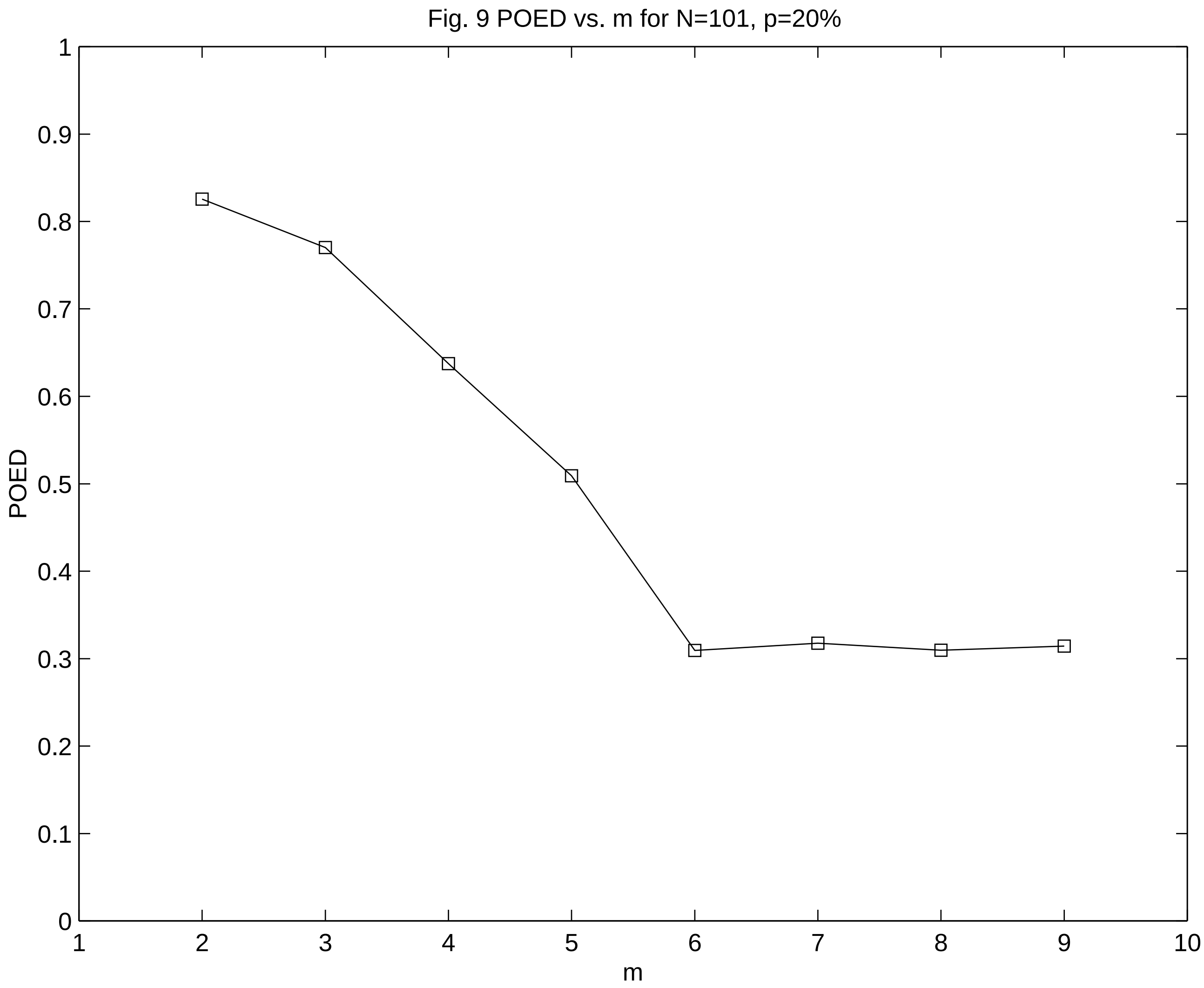
Fig. 9 POED vs. m for N=101, p=20%
POED
m
0
0.1
0.2
0.3
0.4
0.5
0.6
0.7
0.8
0.9
1
1
2
3
4
5
6
7
8
9
10

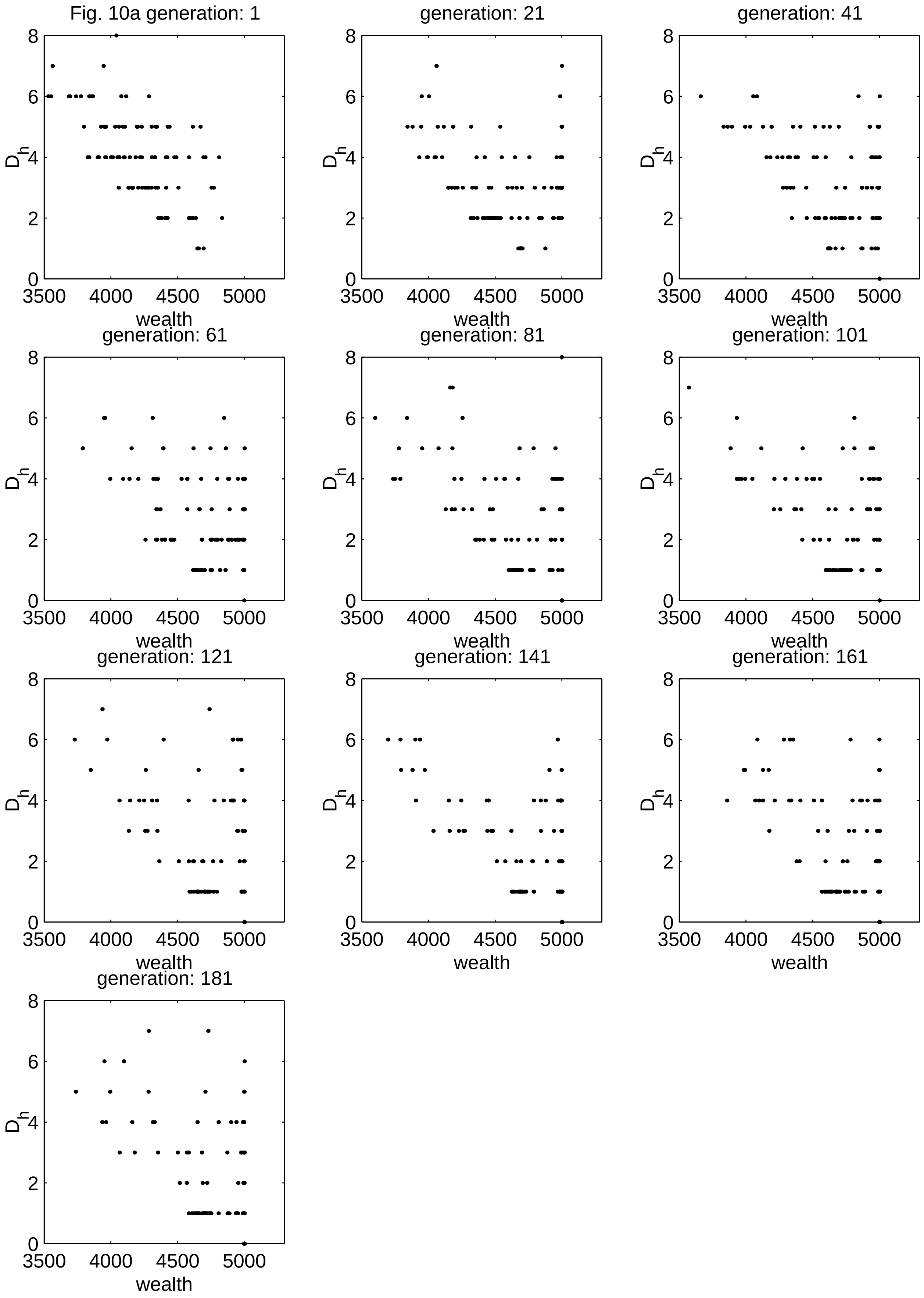

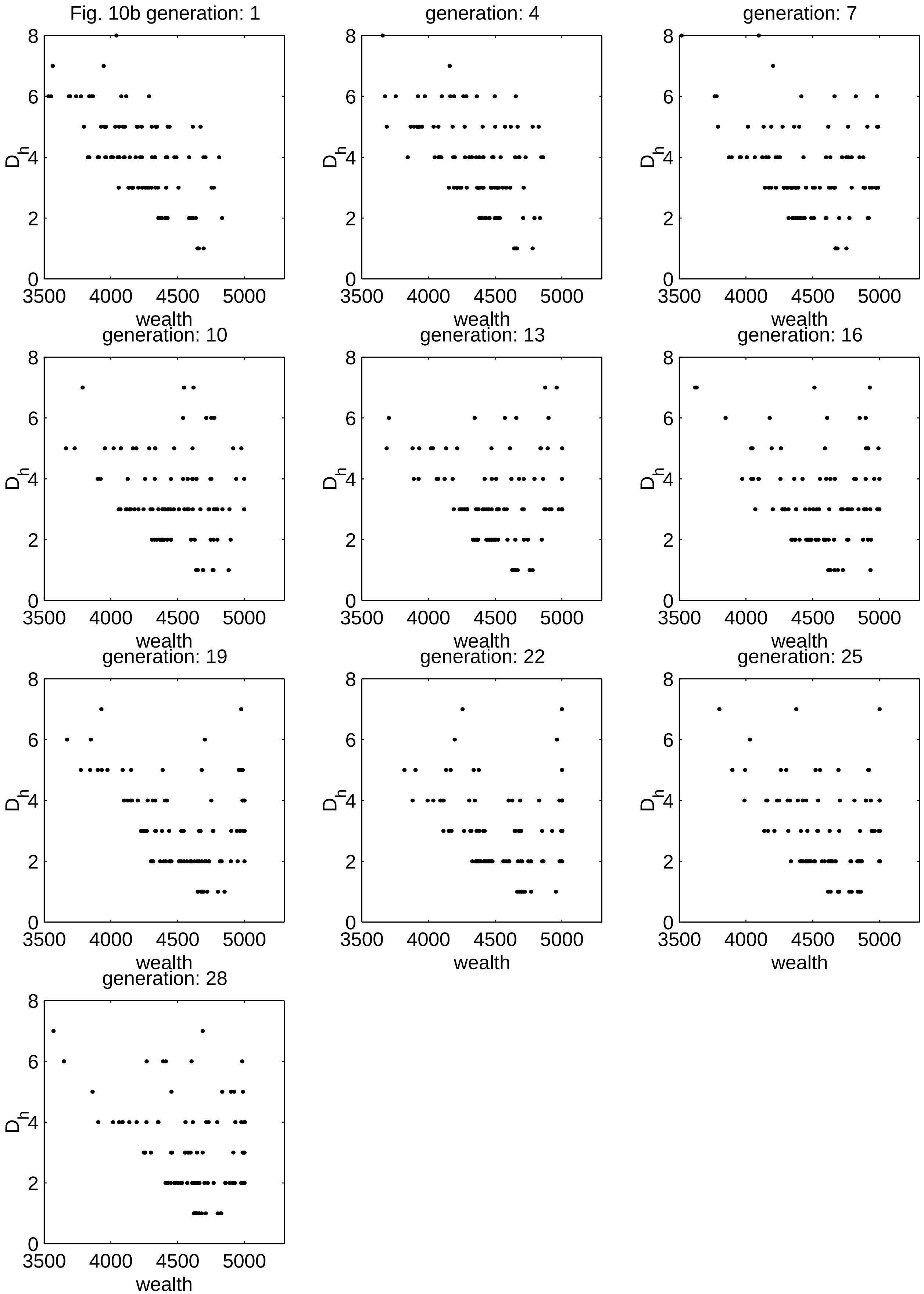

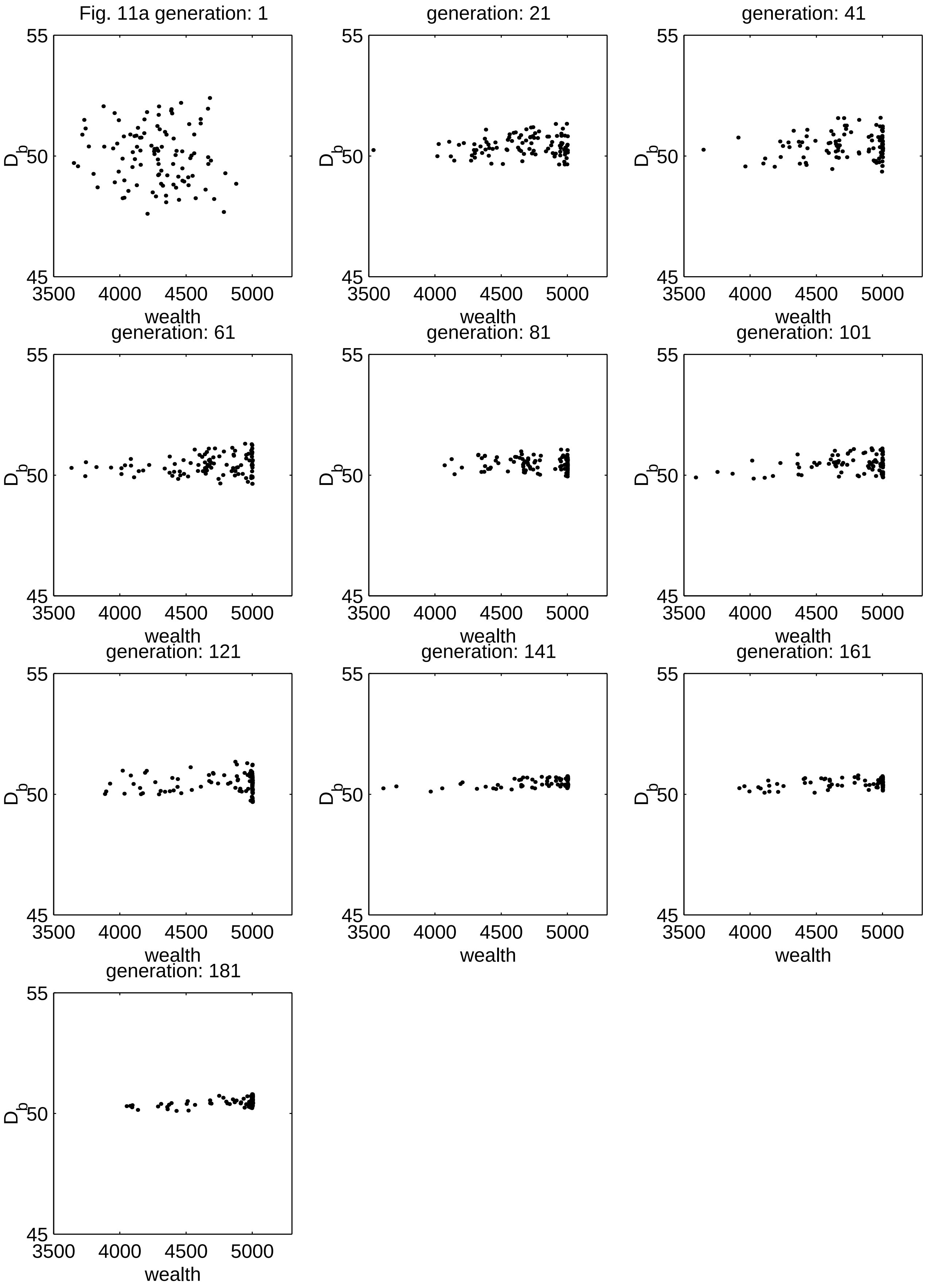

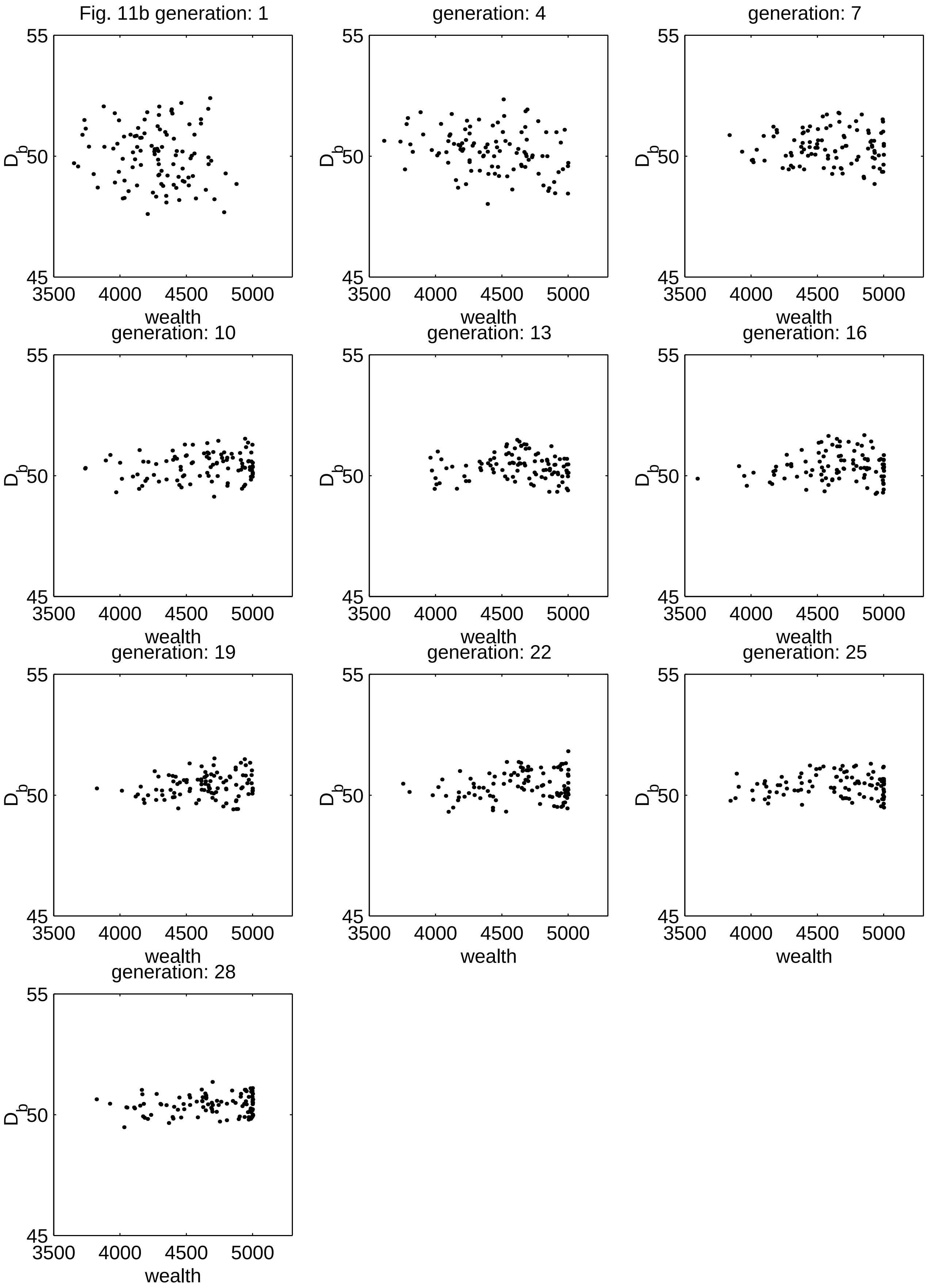

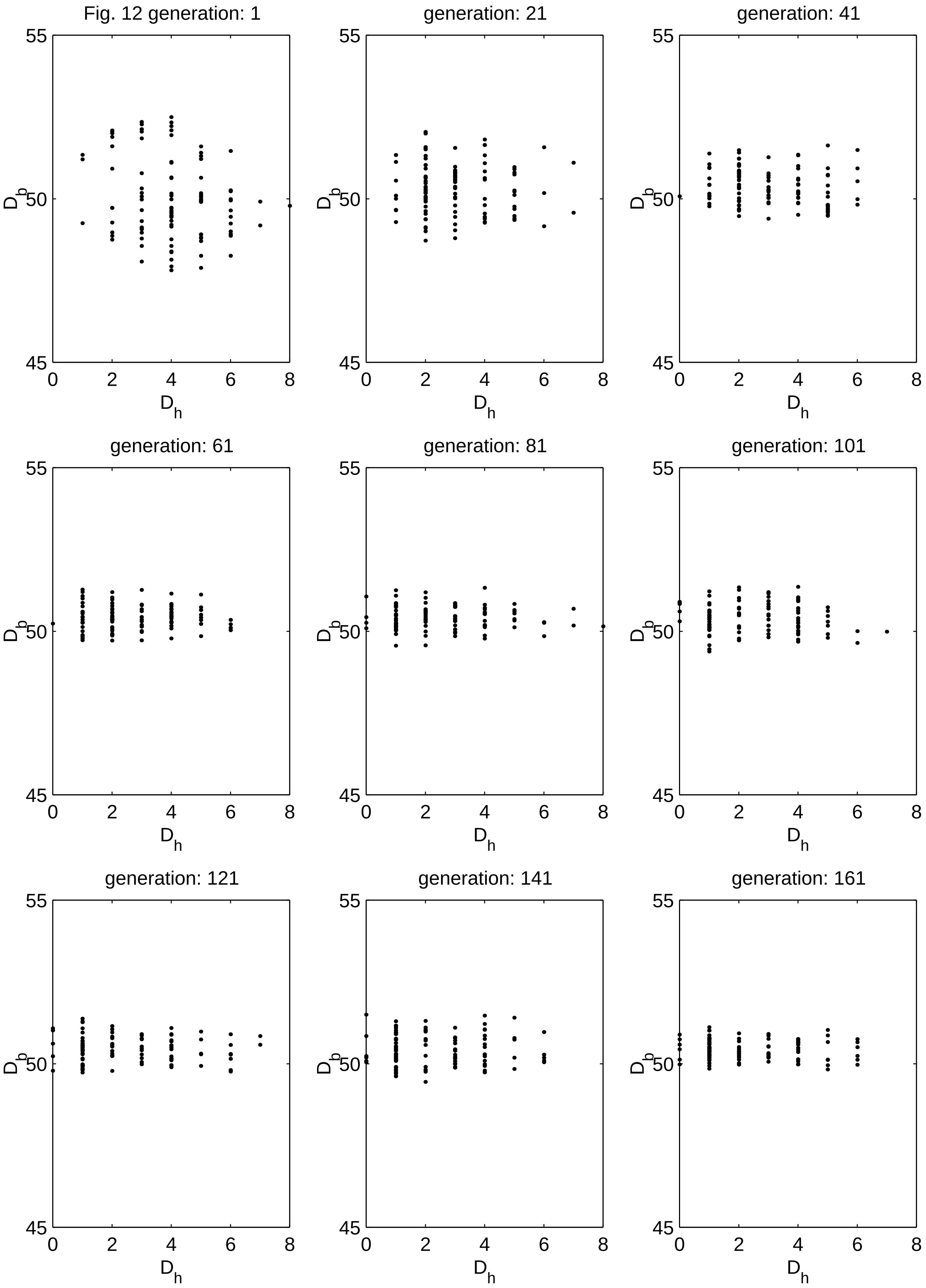

Fig. 13 Standard dev. of $<D_b>$ vs. generation for m=3

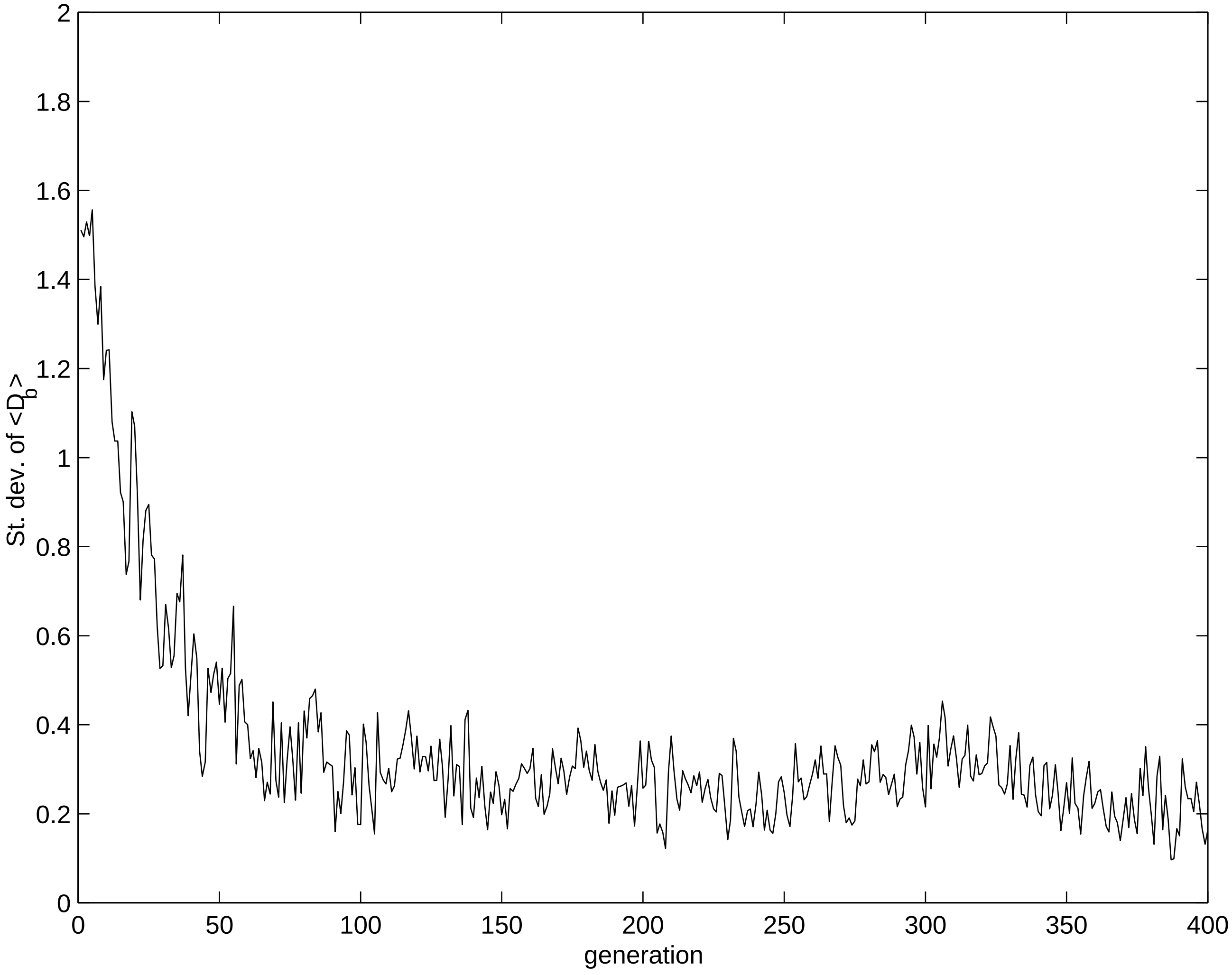

Fig. 14a First generation for N=101 m=3 p=20%

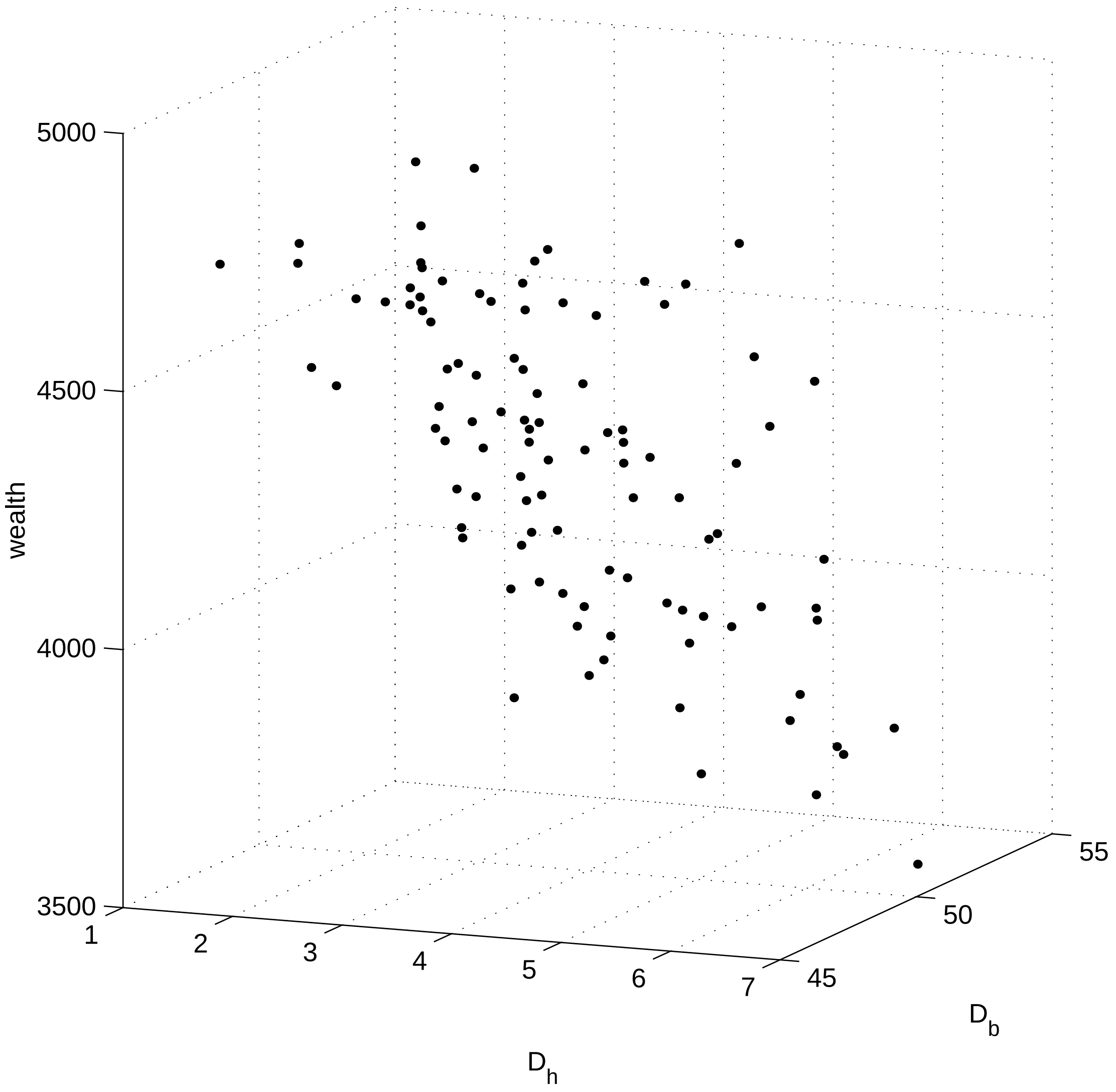

Fig. 14b Last generation for N=101 m=3 p=20%

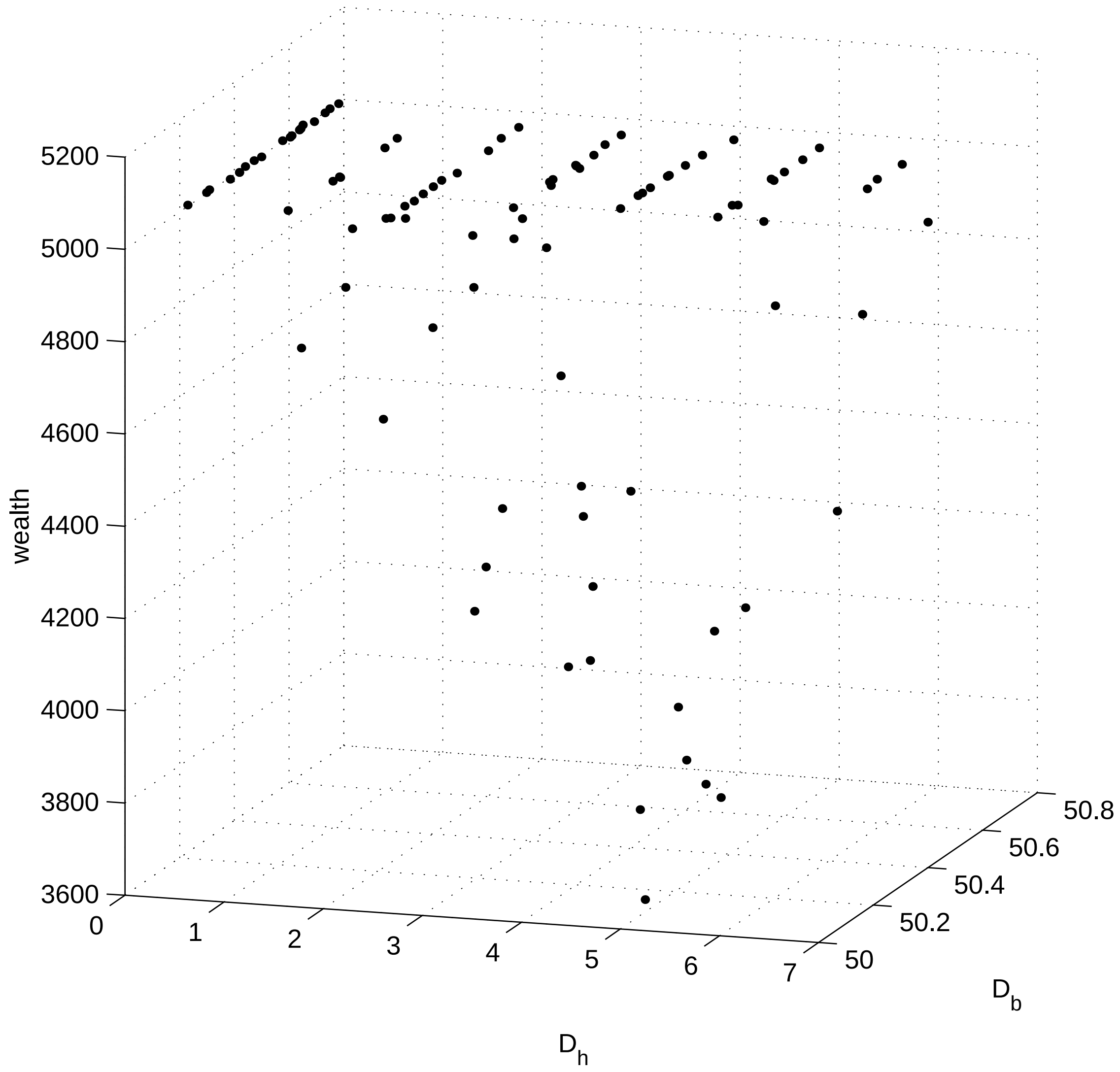

Fig. 15a <D$_h$> vs. generation for m=7

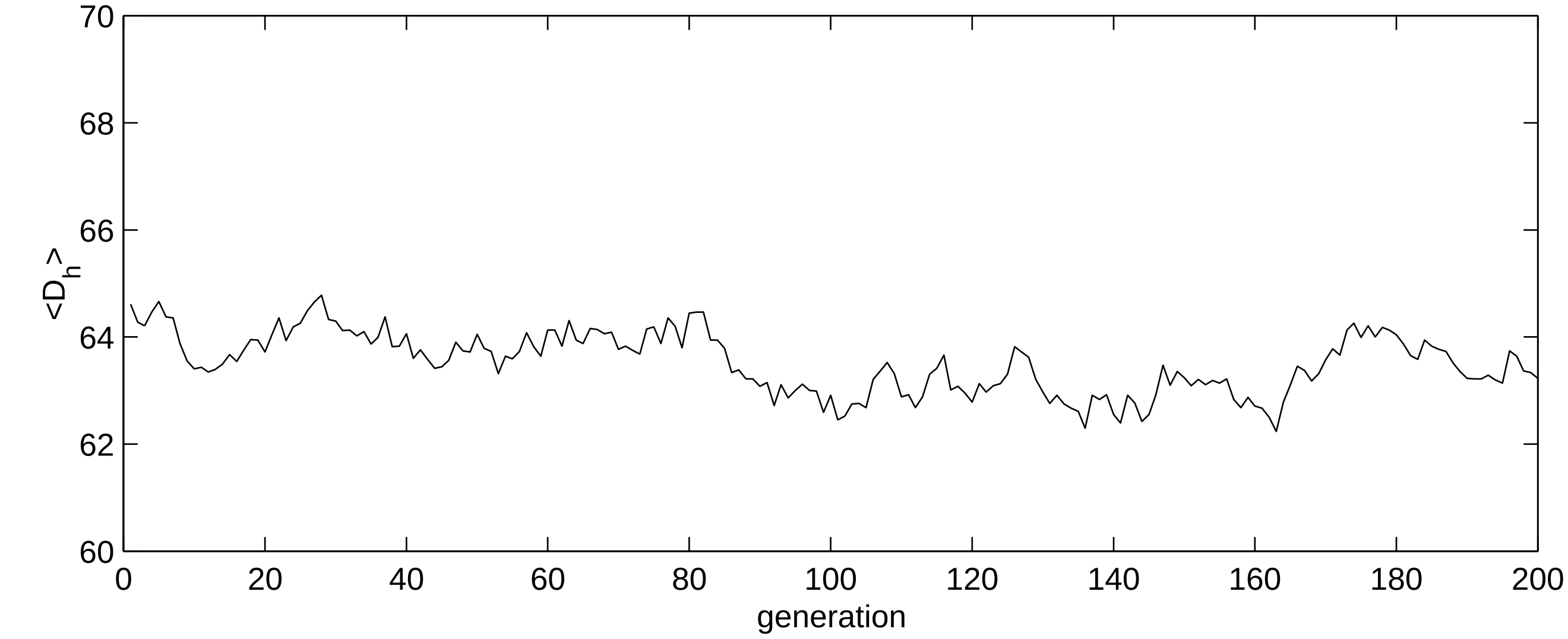

Fig. 15b <D$_b$> vs. generation for m=7

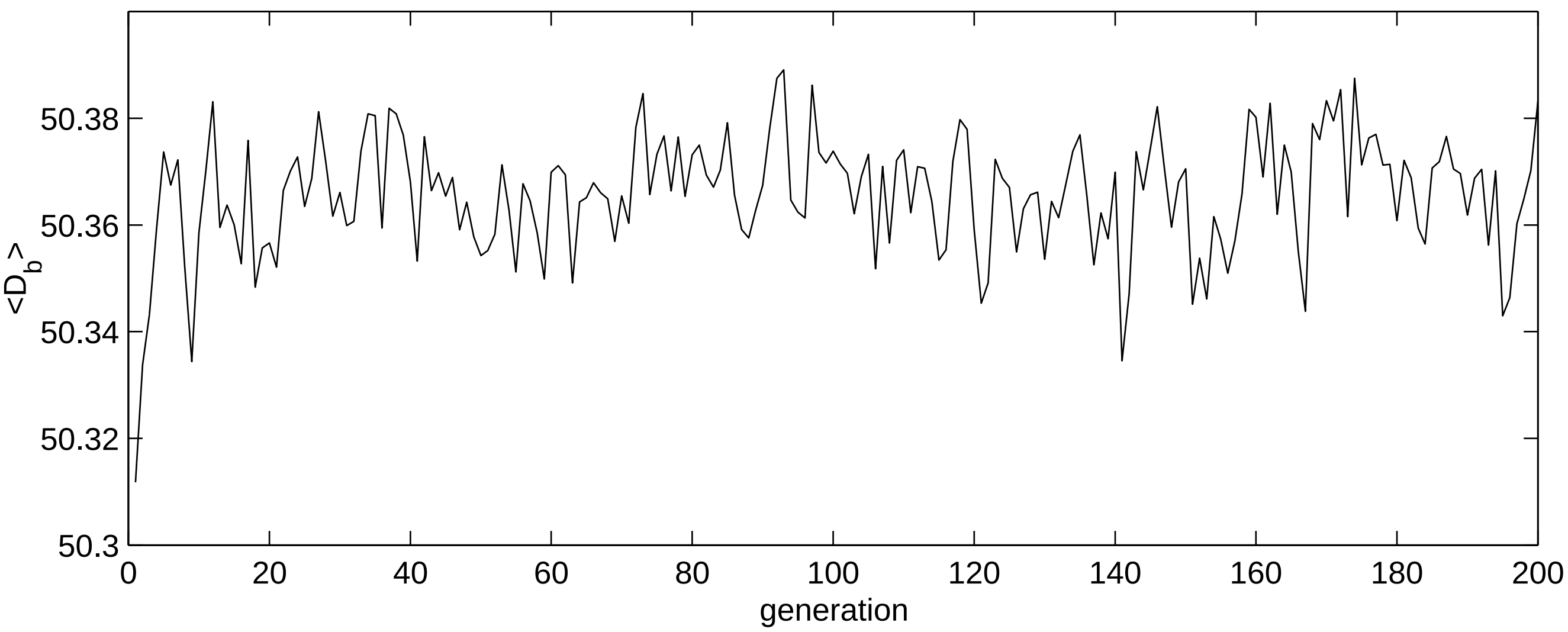

Fig. 15c $\sigma^2$/N vs. generation for m=7

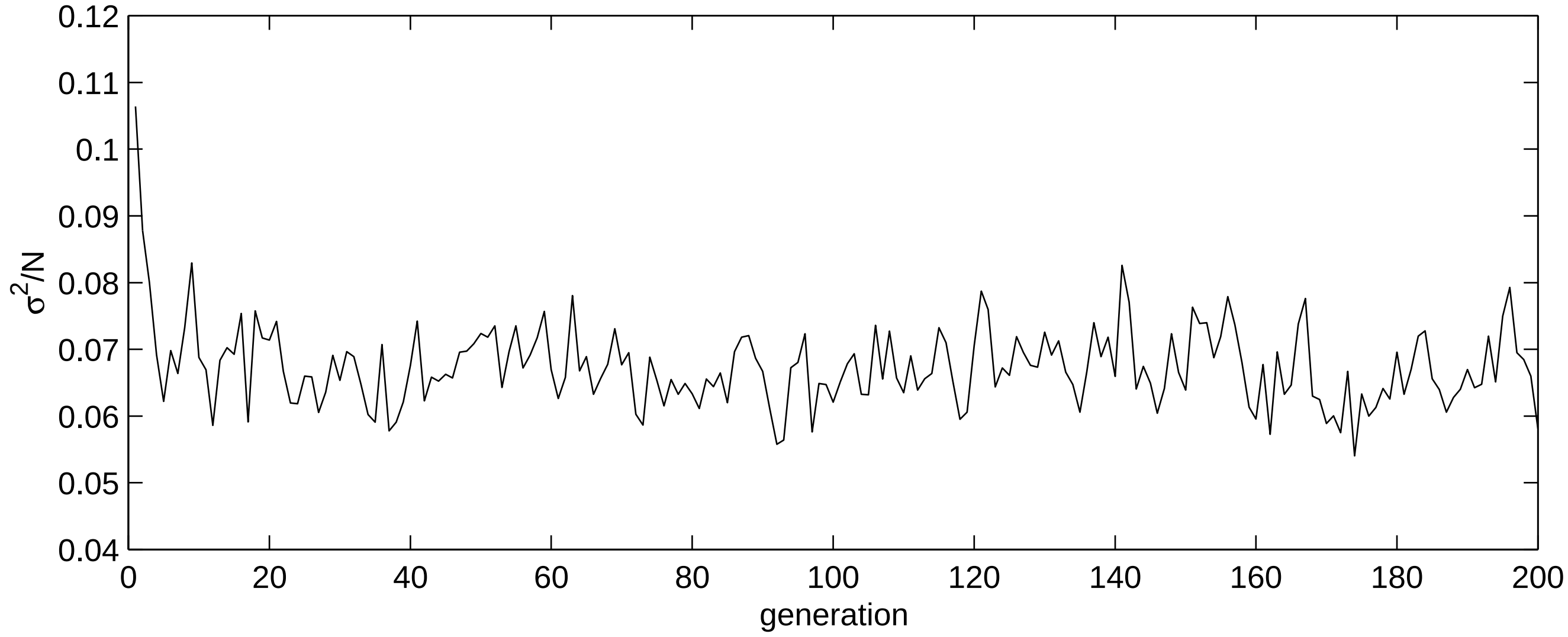

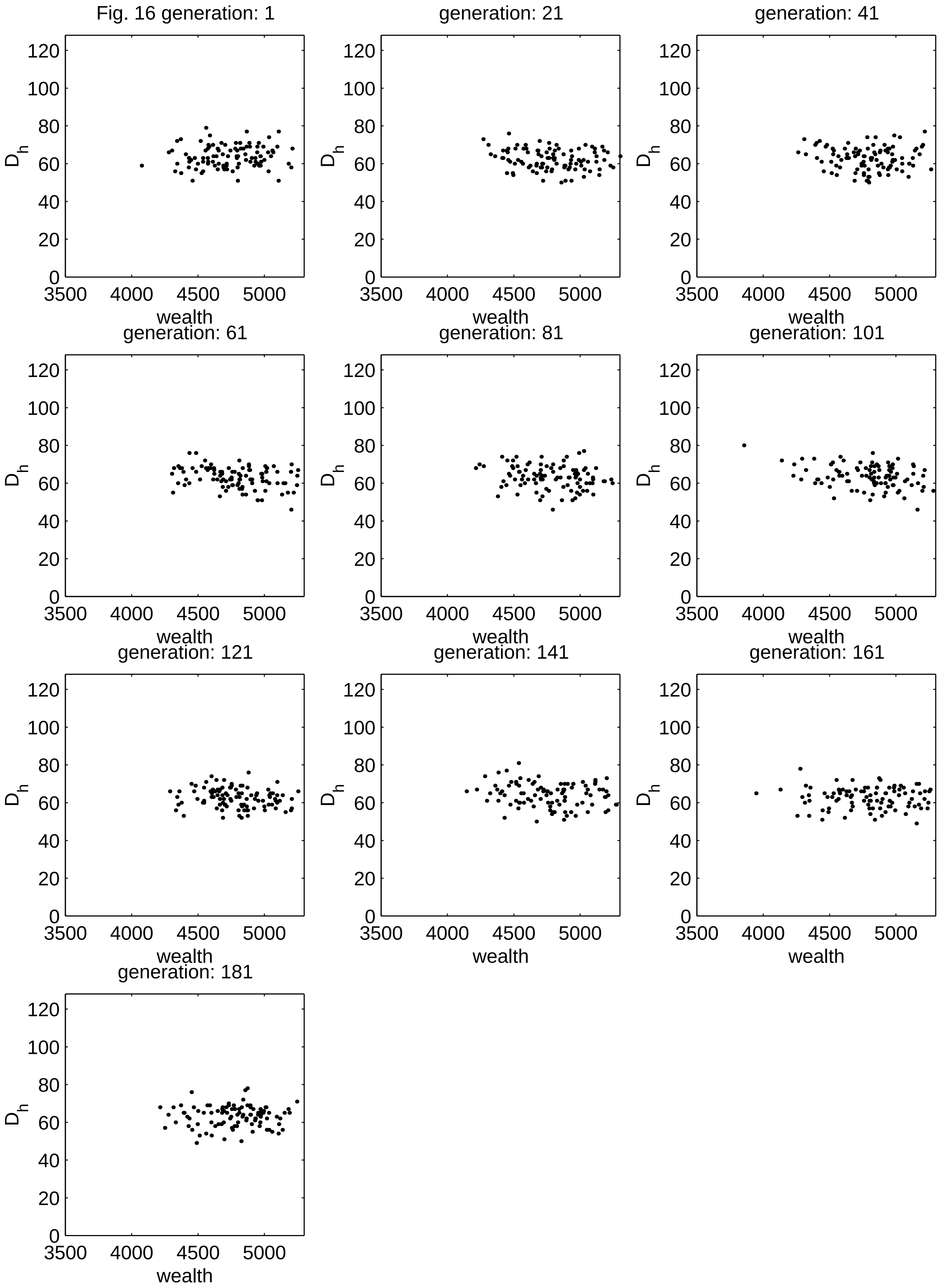

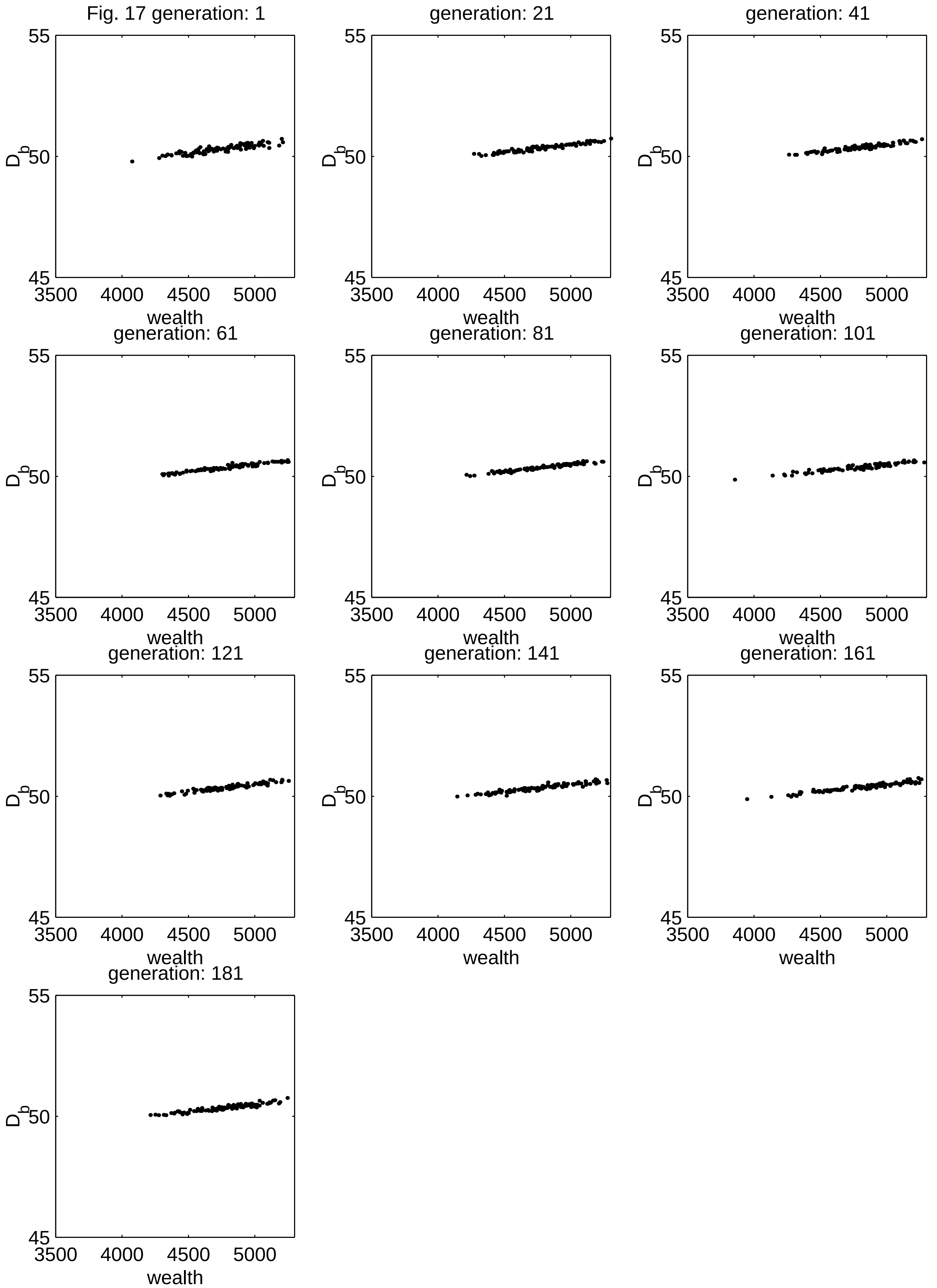

# Evolution in Minority Games
# I. Games with a Fixed Strategy Space

Yi Li, Rick Riolo and Robert Savit
Program for the Study of Complex Systems and Physics Department
University of Michigan, Ann Arbor, MI 48109

Abstract

In this paper we study the minority game in the presence of evolution. In particular, we examine the behavior in games in which the dimension of the strategy space, m, is the same for all agents and fixed for all time. We find that for all values of m, not too large, evolution results in a substantial improvement in overall system performance. We also show that after evolution, results obey a scaling relation among games played with different values of m and different numbers of agents, analogous to that found in the non-evolutionary, adaptive games. Best system performance still occurs, for a given number of agents, at $m_c$, the same value of the dimension of the strategy space as in the non-evolutionary case, but system performance is now nearly an order of magnitude better than the non-evolutionary result. For $m<m_c$, the system evolves to states in which average agent wealth is better than in the random choice game, despite (and in some sense because of) the persistence of maladaptive behavior by some agents. As m gets large, overall systems performance approaches that of the random choice game.

**I. Introduction and Background**

In many biological and social systems agents compete for limited resources. In such systems, it is often the case that the most successful agents are those which act in ways that are distinct from their competitors. Thus, firms which bring new innovations to the market before their competitors are often rewarded, commuters traveling to work at times when the roads are not crowded spend less time and emotional energy traveling, and foraging animals who find an uncrowded plot of land are rewarded by easier access to more food.

One attempt to understand the general underlying dynamics of systems in which agents seek to be different has focused on the analysis of a class of simple games which have come to be known as "minority games".[1,2,3,4] In the simplest version of these games, agents use heterogeneous sets of strategies (in general, different strategies for different agents) to choose, at each time step of the game, to join one of two groups (labeled, say, by 0 or 1). Agents are rewarded if they are in the minority group at a given time step. The most fully studied versions of these games have been adaptive, in that each agent can choose to play a different strategy from his assigned set of strategies at different times in the game. But these games have not been evolutionary, since an agent's individual set of strategies is fixed for the entire game. Evolution, in the broad sense of the appearance of new strategies, is, however, seminally important in the dynamics of complex adaptive systems. In this paper we will include evolutionary effects, and study games in which those agents that perform poorly can try new strategies.

To begin, we will first summarize the structure and basic results of the adaptive, non-evolutionary minority game. In these games, the agents make their choice (to join group 0 or group 1) by following the prediction of a strategy. Strategies make their predictions by using information drawn from a set of common, publicly available information provided to all the agents at each time step. In the simplest case, those data may be drawn from a single time series. For example, one commonly used set of publicly available information (and the one that is used in the cases reported in this paper) is the list of which were the minority groups for the most recent past m time steps. Thus, a strategy is a look-up table with 2 columns and a number of rows. The left hand column contains a list of all possible common signals that the strategy can receive at a given time step of the game. For each such signal, the right hand column contains a 0 or 1 which is that strategy's prediction of which will be the minority group in response to the given signal. For the case in which the strategies use the most recent m minority groups as signals, each strategy table contains $2^m$ rows, corresponding to the $2^m$ possible sequences of m 0's and 1's.

At the beginning of the game, each agent is randomly assigned s such strategies (in general, different, random sets of strategies for different agents). At each time step of the game, an agent must choose which of his s strategies to use. In the simplest versions of the game, each agent, at each time step computes how well each of his s strategies would have done at predicting the correct minority group for all times from the beginning of the game. He then chooses to use that strategy that is currently doing the best. Ties among strategies may be broken in a variety of ways, the simplest being a random choice among the tied strategies. The most intensively studied version of the game is one in which the agents maintain the same strategies during the entire game. Since each agent can choose from among his s strategies the game is adaptive. But the game is not evolutionary, since the strategies are fixed for the duration of the game.

These adaptive, non-evolutionary games, in which all agents have strategies with the same value of m has been studied by several groups[2,3,4,5,6], and the general structure of the game under these conditions is fairly well understood. Such games show a remarkable phase structure in which there is emergent coordination among the agents for a range of values of m. The system-wide behavior can be summarized by considering, $\sigma$, the standard deviation of the number of agents belonging to group 1. The smaller $\sigma$, the larger a typical minority group will be, and thus, the more points will be awarded to the agents in toto. In Fig. 1 we plot $\sigma^2/N$ as a function of $z \equiv 2^m/N$ on a log-log scale for various N and m with s=2. (Scaling curves also exist for other values of s. They are similar in structure to that shown in this figure, but differ in some details.[3]) We see first that all the data fall on a universal curve. The minimum of this curve is near $2^{m_c}/N \equiv z_c \cong 0.5$, and separates two different phases[7]. For $z<z_c$, the system is in a maladaptive phase in which there is no information available to the agents' strategies that can help them predict which will be the next minority group. All the information has been traded away. We call this phase "strategy efficient". The consequence of this efficiency is that the agents' choices tend to be maladaptive, so that the system-wide performance is very poor. For $z \geq z_c$, there is information available to the agents' strategies, and we see an emergent coordination among the agents' choices which results in improved system-wide utilization of resources. The best emergent coordination occurs at $z=z_c$ when the dimension of the strategy space from which the agents draw their strategies is on the order of the number of agents playing the game. As z increases beyond $z_c$, (e.g. as m increases beyond $m_c$ for fixed N), system-wide performance degrades and $\sigma$ approaches the value it would have in the random choice game (RCG), in which agents randomly and independently choose group 0 or group 1 with equal probability. A full description of the dynamics of the non-evolutionary game can be found in Ref. 3.

When one introduces evolutionary effects, thereby allowing the agents' strategies to change as the game proceeds, the results change substantially. Evolutionary games may be classed into two categories: 1. Those games in which all agents respond to the same aspects of the publicly available information, ie., to the same set of signals. In the context of the games discussed here, this amounts to a game in which all strategies in play have the same value of m. Agents may alter their strategies under selective pressure, but may not change the m value of their strategies. 2. Games in which different agents respond to different aspects of the publicly available information. In the context of the games discussed here, this can be most easily implemented by considering games in which different agents have strategies with different values of m. In such games agents may or may not be allowed to change the m-value of their strategies under selective pressure.

In this paper we will discuss the first case. The second case will be discussed in a companion paper.[8] In that work we will see that allowing agents to change the m value of their strategies introduces an interesting, somewhat counter-intuitive new twist to the system. In particular, we will show that the system generally evolves to a state characterized by step function distribution of wealth per agent as a function of m, in which the step transition occurs at a value of $m=m_t$. Agents with $m<m_t$ are relatively wealthy, and agents with $m>m_t$ are relatively poor. We will also show that $m_t \approx m_c - 1$.

In the next section we will describe the evolutionary algorithms and the general, system-wide results in the fixed m games.. For all values of m, not too large, we will see that evolution results in a marked improvement over merely adaptive dynamics, although the best utilization of resources is still at $m=m_c$, as in the strictly adaptive case. In Section III we will study the resulting evolutionary dynamics in more detail, and will provide explanations for the general results presented in Section II. We will also show that the evolutionary dynamics is somewhat different for $m>m_c$ and for $m<m_c$. The paper ends with a discussion and summary in Section IV.

## II. Evolutionary Dynamics and General Results

### A. The Evolutionary Algorithm

We now consider the case in which all strategies have the same value of m. As in the non-evolutionary case, the game begins with N agents randomly assigned s strategies each, of memory m. We also create a random initial history of minority groups of length m+1 so that strategies can be initially evaluated.

We now must specify the evolutionary dynamics. There are many different ways to define evolution consistent with the notion of selective pressure. We have chosen to look at several which are associated with removal of poorly performing strategies. We have not incorporated effects such as incremental mutation or reproduction, although that can easily be done in this context. As we shall explain here and in section III, we find that some central features of our results are independent of the details of the evolutionary processes we have studied. We believe that these features may be yet more general.

To evolve our system, we define a time, $\tau$, which is the duration of one generation. During $\tau$ time steps, the agents' strategies do not change. At the end of $\tau$ time steps, we rank the agents by wealth accumulated during that generation (i.e., how many times they have been in the minority group). We define a "poor" agent to be one whose wealth is in the lowest percentile, p, of agent wealth. We call p the "poverty level". We randomly choose half the agents whose wealth ranks in the lowest p percent, and replace their s strategies with s new, randomly chosen strategies. In the games discussed in this section, all strategies, including the new replacement strategies have the same value of m. (In the games discussed in a companion paper[8], we will allow the replacement strategies to have different values of m.) Those agents whose strategies are not replaced, maintain the relative scores of their strategies from one generation to the next. Agents receiving new strategies have the scores of these new strategies initially set to zero. The game is played for an additional $\tau$ time steps, and the evolutionary process is repeated. In most of the results reported in this section, each agent has s=2 strategies, $\tau$ =10,000 time steps, and p is set so that the impoverished group is defined as either the poorest 10%, 20% or 40% of the population. We will also briefly present results for games played with s=1. This allows us to explore the effects of evolution without adaptivity. Using these parameter ranges, we have studied a variety of games with N=101, 201, 401, and 801 agents run for a total of between 200 and 600 generations.

B. General Results

1. s>1

In this paper we will primarily discuss the case s=2. Other values of s>1 are similar, but differ in some details, similar to the non-evolutionary case.[3] In Fig. 2 we present $\sigma^2/N$ as a function of m for games played with N=101 agents and p=20%. For each value of m, eight independent runs were performed. Each generation was 10,000 time steps, and each game was run until $\sigma^2/N$ was sensibly constant up to fluctuations, generally, 200 generations. The reported values of $\sigma^2/N$ represent an average over the final 50 generations of each run. The horizontal dashed line in this figure is the result $\sigma^2/N$ would have for the random choice game (RCG), in which each agent chooses to join group 0 or 1, randomly and independently with equal probability. This figure resembles Fig. 1 in that i.) $\sigma^2/N$ is a minimum around for m near $m_c$ (in this case, near 5), and ii.) $\sigma^2/N$ approaches the RCG as m gets large[9]. In addition, the spread in values of $\sigma^2/N$ for different runs with a given m is noticeably larger for $m<m_c$ than for $m \geq m_c$ similar to the behavior in the non-evolutionary case.[2,3].

However, there are some important differences. First, the values of $\sigma^2/N$ are generally much lower in Fig. 2 than in the non-evolutionary case, Fig. 1. Most strikingly, $\sigma^2/N$ is less than the value for the RCG for $m<m_c$, in marked contrast to the non-evolutionary case. It turns out that, in the low-m phase, evolution is able to provide a pathway to improved system performance, while still maintaining the quality of strategy-efficiency seen in the purely adaptive, non-evolutionary case. We shall describe below how this comes about. The value of $\sigma^2/N$ near $m_c$ is also remarkably small, being about 1/10 of the value of the RCG. For N=101 agents, this value of $\sigma^2/N$ means that, typically, the minority group is 50 agents half the time, and 49 agents the other half. This is clearly quite close to optimal and is achieved by *emergent* control, not by explicit top-down control. Moreover, the result is robust, varying little from one generation to the next, even though 10% of the agents are replaced after each generation. Notice also that the spread in the values of $\sigma^2/N$ for $m=m_c$ is very small, differing among the runs we have performed only in the third decimal place.

These results also have a remarkable scaling property analogous to the scaling results of the adaptive, non-evolutionary case.[2,3] In Fig. 3 we plot $\sigma^2/N$ as a function of $z \equiv 2^m/N$ for a range of values of m and N. In this figure, each point represents an average of $\sigma^2/N$ over 16 runs with the same values m and N. The poverty level used in all these runs is p=20%. For all values of z, the scaling is quite good, although there is some spread in the results for $z<z_c \equiv 2^{m_c}/N$. This is almost certainly a statistical effect, and follows from the fact that the spread in $\sigma$ for different runs is relatively large for $z<z_c$, as we saw, for example in Fig. 2.

We have also studied the ways in which evolution proceeds for different values of p. In Fig. 4 we plot $\sigma^2/N$ as a function of m for games played with N=101, s=2 and various values of p. In this figure each point represents an average over 8 runs. For each value of m and p, games were played for a long enough time (generally 200 generations[10]) so that $\sigma^2/N$ reached sensibly asymptotic behavior. We see a systematic trend in which, generally, larger values of p are associated with larger values of $\sigma^2/N$. For small p, there are fewer strategies replaced at each generation, and the evolutionary improvement proceeds more slowly. On the other hand, as p increases, selective pressure becomes more indiscriminate, limiting the extent to which the system can improve coordination, leading to a larger asymptotic value of $\sigma^2/N$. In Fig. 5 we plot $\sigma^2/N$ as a function of generation for different values of p, and for m=3 and m=7. In Fig. 5a, we see most clearly a slower initial fall-off for small p, but an asymptotically lower value of $\sigma^2/N$.

An apparent exception to this behavior is at m=5 in Fig. 4. Here it appears that $\sigma^2/N$ is very nearly independent of p. We also note that the difference in the values of $\sigma^2/N$ as a function of p decreases as m=5 is approached from both above and below. As discussed in reference 3, the value of $m_c$ for N=101 is about 5.2. We speculate that at $m_c$, $\sigma^2/N$ is asymptotically independent of p for all $0<p<p^*$, and furthermore, that p* may be one. The detailed nature of the evolved coordinated state at $m=m_c$ that could give rise to this universality is unclear to us, but certainly bears further investigation.

2. s=1

It is also interesting to consider the case in which each agent has only s=1 strategy. This is the situation of evolution without adaptation. We have found that with s=1 there is no significant change in system behavior as a result of evolution. To see this, refer to Fig. 6 in which we plot results for a set of games played with s=1, N=101, p=20%, and various values of m. For each value of m, eight independent runs were performed. In Fig. 6a we plot $\sigma^2/N$ as a function of m for the first generation (of 10,000 time steps). Fig. 6b shows $\sigma^2/N$ for the 200th generation as a function of m. It is clear that there is no systematic difference between the performance of the system with and without evolution. In these games, evolution introduces new and interesting dynamics which can have a significant effect on the performance of a system, but only if the agents are also adaptive. Replacing poorly performing random strategies by other random strategies does not lead to real selective pressure, unless there is some additional intra-agent dynamics.

## III. Understanding the General Results

We now want to try to understand the dynamics that gives rise to some of these general results. To do so, it will be important to introduce two distance measures associated with properties of individual agents.[3] One is an intra-agent distance, $D_h(i)$, defined as the Hamming distance between the $i^{th}$ agent's two strategies. The second is a distance in "behavior space", $D_b(i)$, and may be understood to be the average behavioral distance of the $i^{th}$ agent from all other agents playing the game. In particular, let $T_i^{[j]}(u_m)$ denote the response (0 or 1) to the string, $u_m$, of the $j^{th}$ strategy of agent i, and let $\phi_i(j)$ denote the probability that agent i uses strategy j. Further, let $P(u_m)$ be the probability that the m-string $u_m$ appears in the sequence of minority groups. Then

$$D_h(i) = \sum_{u_m} \left| T_i^{[1]}(u_m) - T_i^{[2]}(u_m) \right| \qquad (2.1)$$

and

$$D_b(i) = \sum_j \sum_{k,l} \sum_{u_m} P(u_m)\phi_i(k)\phi_j(l)\left|T_i^{[k]}(u_m) - T_j^{[l]}(u_m)\right|. \qquad (2.2)$$

A. The Low-m Phase

Let us first consider evolutionary dynamics in the low-m phase, $m<m_c$. As described in Ref. 3, in the purely adaptive, non-evolutionary case, the system manifests maladaptive behavior in this phase. Typically, odd occurrences of a given m-string result in more or less, random choices by the agents (giving rise to a minority group with a population close to 50%, within random fluctuations). However, even occurrences of a given m-string give rise to very small minority groups: In this case, agents use information about the group's last response to a given string and exhibit a herding behavior, in which many agents join the opposite group. Although in this phase, no agent ever earns more than 50% of the possible points, those that do the best tend to have strategies whose relative Hamming distances are relatively small. Small Hamming distances means that the agent's strategies are relatively similar. Thus the agent is often prevented from being able to make a maladaptive choice. The worst performing agents, on the other hand, have strategies whose Hamming distances are large, thus allowing them to (maladaptively) "follow the crowd" and join the majority group much of the time.

Although the best predictor of agent wealth in the non-evolutionary game is Hamming distance (the smaller the better), we have found that when these systems are allowed to evolve, evolutionary dynamics selects for two different traits. Wealthy agents turn out to be those with either small values of $D_h(i)$ or large values of $D_b(i)$. Moreover, in a low-m game, evolution proceeds in two moderately distinct stages. First, since the poorly performing agents are preferentially removed from the system, one would expect evolution to lower the average Hamming distance between the agents' strategies. Indeed, this is what we see. In Fig. 7a we plot the average Hamming distance between the agents' two strategies as a function of generation for a game played with m=3 and N=101. (Results for other values of $m<m_c$ are qualitatively similar.) We see a very clear, rapid drop off of the average Hamming distance in the early stages of evolution, up to about 40 generations. There is also an improvement of overall resource utilization, as can be seen in Fig. 7c in which we plot $\sigma^2/N$ as a function of generation for the same run. At the same time there is also a relatively rapid increase in $\langle D_b \rangle$ as can be seen in Fig. 7b, suggesting that agents with small $D_b(i)$ are also selected against, even in this early stage of evolution. In this example, the early stage persists for the first 40 or so generations. Following this first stage of evolution, a second stage sets in, in which $\langle D_h \rangle$ fluctuates without falling much further, and $\langle D_b \rangle$ continues to rise slowly. By generation 100 or so, both $\langle D_b \rangle$ and $\sigma^2/N$ have reached asymptotic values, within fluctuations. The cross-over between these two stages of evolution (in this example, at about generation 40) occurs when $\sigma^2/N$ (Fig. 7c) is close to about 0.25, the value found in the RCG. Other runs performed with $m<m_c$ generally show evidence of this two stage evolutionary structure, although not always as clearly as the example in Fig. 7.

1. Period-two dynamics and the role of $D_h(i)$.

To understand what's going on, look at Fig. 8, in which we plot the conditional probability $P(1|u_m)$, for 1 to be the minority group following a specific string of length m for the game played with m=3, N=101. Fig. 8a shows $P(1|u_m)$ for the first generation, and Fig. 8b shows $P(1|u_m)$ for the last generation (in theses runs, the 400$^{th}$ generation). That the histogram in Fig. 8a is flat is what we expect[3], but what is remarkable is that the histogram late in evolution is also very nearly flat. That is, the system in the low-m phase continues to be very nearly (but not entirely) strategy-efficient, even after evolution, but at the same time shows good system-wide performance in that $\sigma^2/N$ is much smaller than in the RCG. How does this come about?

Recall that the flat histogram in the non-evolutionary case is due to an embedded period-two dynamics in which even occurrences of a given string result in very small minority groups. Even after evolution, the system in the low-m phase possesses the same period two dynamics, but the consequence of the maladaptive, herding behavior is less dramatic, and in fact, and somewhat ironically, leads to system-wide performance better than that of the RCG. To see that period-two dynamics still dominates the low-m phase, refer to Fig. 9. Here we plot POED, the probability that the minority group in response to an even occurrence of a string is different than the minority group following the preceding odd occurrence of the same string, as a function of m, for games played with N=101 agents. We see that for $m<5\approx m_c$, POED is significantly greater than ½, indicating the presence of significant period-two dynamics.

To understand the consequence of period-two dynamics in an evolutionary context, consider the example run referred to in Figs. 7 and 8, with m=3, N=101. Note, first, from Fig. 7a, that late in the evolution, the average Hamming distance has dropped to about 2.75.[11] This means that, on average, an agent's two strategies differ in their responses to 2.75 out of 8 possible strings. Therefore, typically, 5.25/8 (=65.63%) of agents must always respond to occurrences of a given string (call it $u_m$) in the same fixed way, since both their strategies will dictate the same response. On the other hand, if the period-two dynamics still obtains, then the remaining 34.37% of agents will be able to either choose randomly between the two groups in response to an odd occurrence of $u_m$, or adapt (and, as in the non-evolutionary case, *mal*-adapt) to an even occurrence of $u_m$. Now, of the 66 or so agents whose strategies dictate the same response to $u_m$, typically, about 29 (=1/2[66–$(66)^{1/2}$]) will always join one group (say, group 0), while the remaining 37 will always join the other group (say, group 1). Of the remaining 34 agents, roughly half (17±2) will join each group. Thus, in response to an odd occurrence of $u_m$ in this example, group 0 will almost always be the minority group, since 29 (fixed responses) + 17±2 (adaptive responses)<51.

Next, consider the response to an even occurrence of $u_m$. In this case, as with the odd occurrence, 29 of 66 agents will again join group 0 and 37 will join group 1. Of the remaining 34 agents, roughly 22 will join group zero. The reasoning is as follows: If all of the remaining 34 agents differed only in their response to $u_m$, then, by the usual arguments of period-two dynamics[3], all 34 would join group 0. However, since the average Hamming distance is about 2.75, roughly ¼ of the 34 agents differ in their response to one other string, and roughly ¾ of the 34 agents differ in their response to two other strings. If the even occurrence of $u_m$ happens to lie between an odd and an even occurrence of another string in which the two strategies differ, then there will be roughly a 50% probability that the relative rankings of the two strategies will be changed, in which case, the agent will join group 1 rather than group zero. If an agent's strategies differ in their responses to only one additional string, other than $u_m$, the probability of that agent joining group 1 is about 25%. I.e., the probability of the even occurrence of $u_m$ lying between an odd and even occurrence of the other string is about 0.5, and if that happens, the *a priori* probability of the rankings of the two strategies being altered is also 0.5. A similar argument for the case in which an agent's two strategies differ in their response to two strings aside from $u_m$, shows that the probability of that agent joining group 1 is about 37.5%. Thus, out of 34 agents, about 12 will join group 1, leaving 22 to join group 0. Consequently, in response to an even occurrence of $u_m$, about 51 agents will join group 0, leaving 50 in the minority group.

In this example, then, we see that in response to an odd occurrence of a given m-string, the system will almost always choose group 0 as the minority group, but with a minority group population of about 45 (consistent with a typical random result). But in response to even occurrences of a given m-string, the minority group population will be usually be about 50, which is nearly optimal, and is significantly better than random. This leads, on average to a value of $\sigma^2/N$ of about 0.1. this is roughly consistent with the value of $\sigma^2/N$ at the end of the first stage of evolution (about generation 40), and within a factor of two (but see the next paragraph) of the result observed late in the evolution of the m=3 games with 101 agents (Fig. 7b). Note that although there is marked improvement over the system-wide results of the RCG, the dominant dynamics of the system in the low m-phase is still that of period-two dynamics. Moreover, those agents that, in response to an even occurrence of a given m-string, join the group that was the minority group following the previous odd occurrence of that same string are still behaving maladaptively, and are still losing points. This is borne out by Fig. 10a, in which we plot agent wealth versus $D_h(i)$ every 20 generations for this example, and in Fig. 10b in which we plot the same quantity every third generation for the first 30 generations. We see that, even after evolution, the wealthy agents tend to have small values of $D_h(i)$. And it is because of the diminution of the average Hamming distance forced by evolution that the system is able to limit the number of agents who make maladaptive choices. *The irony is, that it is precisely during those times (i.e., in response to even occurrences of m-strings) when a limited number of agents make maladaptive choices, that the typical population of the minority group comes closer to 50% of the agents, lowering the average value of $\sigma^2/N$, and resulting in an improvement in the general good.*

2. <u>The evolutionary role of $D_b(i)$.</u>

Although the most important dynamic driving the evolutionary improvement in $\sigma^2/N$ for $m<m_c$ is bound up with a decrease in $\langle D_h \rangle$ and the role of the period-two dynamics, that is not the whole story. The argument in the last paragraph leads to an expectation that, for m=3, N=101, $\sigma^2/N$ should be about 0.1, late in evolution. But as we see from Fig. 7b, $\sigma^2/N$ is about half that. The remaining improvement in $\sigma^2/N$ is related to the fact that evolution also selects against agents with small $D_b$. The fact that there is an increase in $\langle D_b \rangle$, relatively rapid in the early stage of evolution, and slower in the late stage, suggests this.

To explore this a little more fully, refer to Fig. 11, in which we show a series of scatter plots of agent wealth versus $D_b(i)$ for series of generations during evolution. Fig. 11a shows the scatter plots every 20 generations for 200, and Fig. 11b shows the scatter generations plots every three generations for the first 30 generations. Note first that in the early stages of evolution there a large spread in $D_b(i)$. This spread decreases during the first stage of evolution, and by generation 40 the distribution in these plots has narrowed considerably. By this point, most agents have values of $D_b(i)$ between 49.5 and 51. As evolution continues, there is an increasingly strong and clear correlation between an agent's wealth and his value of $D_b(i)$[12]. The large initial spread in $D_b(i)$ is due to the small size of the available strategy space (small m), as we shall explain below. The rapid decrease in the spread of $D_b(i)$ during the first stage of evolution is due, largely, to selection against agents with large values of $D_h(i)$ and in part to selection against agents with small values of $D_b(i)$, as we shall now explain.

First, to understand the origin of the large spread in $D_b(i)$, refer to Fig. 12, in which we plot $D_b(i)$ as a function of $D_h(i)$ every 20 generation for a game played with m=3. We see that early in the evolution, agents with either very large or very small values of $D_b(i)$, generally have values of $D_h(i)$ close to 4 (i.e. close to $2^{m-1}$). To understand this, suppose an agent's two strategies differ in c of the $2^m$ entries in his two strategies, so that $D_h(i)$ is c. Now let's estimate $D_b(i)$ for this agent. Consider one of the c $u_m$'s for which the agent's strategies differ. For simplicity, let us set $\phi_i(k)$ and $\phi_j(i)$ both equal to ½. The contribution from such a $u_m$ to the sum in eq. (2.2) over k, l and j, will be exactly $P(u_m)(N-1)/2$, or, in our example, $50P(u_m)$. This is because each $T^{[l]}_j$ will differ from either $T^{[1]}_i$ or $T^{[2]}_i$, but not both. Thus, an agent whose Hamming distance is $2^m$ will have a value of $D_b(i)$ of exactly (N-1)/2. If $c<2^m$, however, $D_b(i)$ will differ from 50. Now, if both an agent's strategies have the same response to a given $u_m$, then the contribution of that string to $D_b(i)$ will depend on the specific distribution of 0's and 1's in the responses of the other agents' strategies to that string. If an agent has a relatively low value of $D_h(i)$, then there will be many such strings contributing to $D_b(i)$. The mean of such contributions, averaged over many strings will be about 50, and if there are many such strings, then we expect that the relative deviation from 50 will be fairly small. Indeed, this is what we see in games played in the high m-phase,[3] in which c is typically of order $2^{m-1}$, and in which it is very rare for c to be close to $2^m$. On the other hand, if c differs from $2^m$ by only a few, which is not uncommon for small m, (and means that the agent has a relatively large Hamming distance), then $D_b(i)$ will be very sensitive to the fluctuations in the distribution of 0's and 1's to all agents' strategies in response to a few m-strings. In this case, the fluctuations in $D_b(i)$ about 50 may be relatively large. For the game played with m=3 and N=101, a simple estimate shows that for agents with Hamming distances of 7, we should expect values of $D_b(i)$ in the range of ~49.5 to ~50.5 (50±½), and for agents with Hamming distances of 4, expect values of $D_b(i)$ in the range of ~48 to ~52 (50±2).

Selection against large values of $D_h(i)$ also induces a narrowing in the spread of the $D_b(i)$ distribution. In Fig. 7a, we see that, in this example, $\langle D_h \rangle$ falls from about 4 to about 2.75 during the first stage of evolution. But in Fig. 12, we saw that values of $D_h(i)$ near 4 are associated with the extreme values of $D_b(i)$. Thus, as more agents are driven to smaller values of $D_h(i)$, the spread in the values of $D_b(i)$ also narrows.

Although the most important dynamic in the early stage of evolution is selection against large $D_h(i)$, there is also selection against small $D_b(i)$. We note that these are not the same effects since, as we see from the first plot in Fig. 12, large values of $D_h(i)$ are associated, fairly symmetrically, with both small and large values of $D_b(i)$. Thus, if the narrowing of the distribution of $D_b(i)$ were due entirely to selection against high values of $D_h(i)$, we should expect that narrowing to occur fairly symmetrically. However, this is not what happens. Look again at Fig. 11b. We see that during the first 20 or so generations, the distribution of $D_b(i)$ is depleted more readily on the low side than on the high side. Somewhat later, the distribution of values of $D_b(i)$ shrinks also from the high side. Thus, independent selection against small values of $D_b(i)$ also occurs in the early stage of evolution.[13]

In Fig. 13 we plot the standard deviation of $D_b(i)$ as a function of generation for the example we have been discussing, with m=3. This quantity stops its rapid decrease after about 40-50 generations. This is about the same time at which $\langle D_h \rangle$ stops its rapid decrease, and $\langle D_b \rangle$ changes from increasing rapidly to increasing more slowly. This marks the end of the first stage of evolution in games with $m<m_c$. At this point, evolutionary selection against agents with high $D_h(i)$ ceases to be important, the width of the distribution values of $D_b(i)$ has narrowed considerably, and the distribution of agent wealth versus $D_b(i)$ begins to resemble that associated with games played in the high-m phase[3]. Further evolutionary improvement in system performance now relies primarily on selection against agents with low values of $D_b(i)$.

3. Two wealthy groups

As a consequence of the interplay of natural selection with low-m minority dynamics, evolution proceeds, roughly, in two stages, the first selecting against both high values of $D_h(i)$ and low values of $D_b(i)$, and the second further selecting against low values of $D_b(i)$. This produces two groups of wealthy agents with different traits, either low values of $D_h(i)$ or high values of $D_b(i)$ when $m<m_c$. We can see this directly, by referring to Fig. 14. Here we present three-dimensional scatter plots of agent wealth versus $D_b(i)$ and $D_h(i)$, after the first and the last generation for a typical run with m=3, N=101. Note, in particular, that late in evolution wealthy agents may have either a small value of $D_h(i)$ or a large value of $D_b(i)$, but not necessarily both. Those with small Hamming distance take advantage of the maladaptive behavior of a relatively small number (about 20%, in the example above) of agents who continue to drive the period-two dynamics. Of those agents with larger Hamming distance, there is a subset with large distances in behavior space, whose choices are commonly different than the majority of other agents in the game, and whose wealth is correspondingly high. Notice also that this group shows up late in evolution in Fig. 10a. In later generations, there is, in general, a general, strong, inverse relationship between wealth and $D_h$. But there is also a cluster of agents with high wealth and large $D_h$. This group is most evident in the plots of generations 141, 161, and 181. This group does not directly rely on the period-two dynamics for its high wealth. Rather, they are the harbingers of the much larger population of wealthy agents far separated in behavior space that dominate the dynamics for $m \geq m_c$, but not too large.

B. The High-m Phase

Consider now the effects of evolution for $m \geq m_c$. For these values of m, there is no maladaptive behavior in the non-evolutionary game. Rather there is emergent coordination leading to a better than random utilization of resources, even in the absence of evolution. Nevertheless, evolution improves the situation still further. In this case, evolution is a single stage process. Because there is no maladaptive behavior, agents are not selected for on the basis of the Hamming distance between their strategies. Rather, poorly performing agents are those for which $D_b(i)$ is relatively small. In Fig. 15, we plot $\langle D_h \rangle$ and $\langle D_b \rangle$ as a function of generation for a game played with m=7, N=101 and p=20%. Here we see no systematic change in $\langle D_h \rangle$ over time, but we do see a sharp and rapid increase in $\langle D_b \rangle$ over the first five or so generations. We have also plotted in this figure $\sigma^2/N$ as a function of generation. There a sharp and rapid decrease in $\sigma^2/N$ corresponding to the rapid increase in $\langle D_b \rangle$. This supports our picture that for $m \geq m_c$ evolution selects against agents with a small value of $D_b(i)$, forcing a more efficacious distribution of agents in behavior space, and consequently a lower value of $\sigma^2/N$. Finally, refer to Figs. 16 and 17 in which we show scatter plots of agent wealth versus $D_h(i)$ and agent wealth versus $D_b(i)$, respectively, every 20 generations for a run with m=7, N=101 and p=20%. Unlike the corresponding plot for m=3 in Fig. 11, Fig. 16 shows no correlation, even late in the evolution between low values of $D_h(i)$ and agent wealth. On the other hand, the plot of $D_b(i)$ versus agent wealth for this game (Fig. 17) is qualitatively similar to the plot of $D_b(i)$ versus agent wealth for m=3 (Fig. 11) late in evolution, although with less spread. Thus, for $m < m_c$, evolution selects both for low $D_h(i)$ and high $D_b(i)$ reflecting the period-two dynamics, and the sometime maladaptive behavior of various agents in the system. For $m \geq m_c$ the period-two dynamics is not significant, and evolution selects only for large $D_b(i)$.

## IV. Summary and Discussion

Summary

In this paper we have demonstrated the following features of evolution in minority games with fixed strategy space:

1. Evolution results in significant improvements in system-wide performance for all values of m. For $m<m_c$, $\sigma^2/N$ is markedly smaller than in the non-evolutionary game, and is, in fact, less than the RCG value. For $m=m_c$, $\sigma^2/N$ is nearly an order of magnitude smaller than in the non-evolutionary, adaptive version of the game.
2. After evolution, all values of $\sigma^2/N$ for games with different m and N (but with the same value of p) lie on a universal curve as a function of $z=2^m/N$. As in the non-evolutionary case, the minimum of the curve is at $m=m_c$, at which point there is a phase change as a function of m. Also as in the non-evolutionary case, the low-m phase is a (nearly) strategy-efficient phase, characterized by the dominance period-two dynamics. Ironically, after evolution, it is the maladaptive choices of a small percentage of agents that is responsible for most of the improvement in the system-wide behavior. The high-m phase is characterized by emergent coordination among the agents' choices, similar to the non-evolutionary case. Except for very large z, when the behavior of the system is similar to that of the non-evolutionary case and approaches that of the RCG, the $\sigma^2/N$ curve is lower than in the non-evolutionary case.

3. In the low-m phase, evolution can be roughly separated into two stages. In the first stage the dominant dynamics is selection against agents with large values of $D_h(i)$, although there is also selection against agents with small values of $D_b(i)$. The second stage of evolution is dominated by selection against small values of $D_b(i)$. In the high-m phase evolution proceeds in one stage, in which there is selection against small values of $D_b(i)$.
4. The precise values of $\sigma^2/N$ generally depend on p, the parameter in the evolutionary algorithm which determines the fraction of the low performing agents that adopt new strategies in each generation. However, it appears that the result is independent of p at $m=m_c$.
5. For games played with one strategy per agent, evolution does not materially alter the average behavior of the system. In the context of these games, evolution is only effective when the agents are already adaptive.

Discussion

It is not surprising that evolution generally improves system-wide performance. But the degree to which that performance is improved, and the ways in which the dynamics achieve that improvement are surprising. The dynamics in the low-m phase is particularly interesting. It is remarkable that in such a simple system one can identify distinct traits (low $D_h(i)$ and high $D_b(i)$) that are selected for, leading to a heterogeneous population of wealthy agents. And the intricate way in which evolution arranges for good collective performance, even in the presence of period-two dynamics is fascinating.

Equally surprising is the existence of scaling behavior in these systems, even after evolution. While evolution changes the shape of the scaling curve, it does not change the value of $z_c$. Although a full understanding of scaling is still lacking, it is clear that this robust property is deeply embedded in the relationship between the geometry of the strategy space and the adaptive competition. In addition, the apparent p-independence of $\sigma^2/N$ at $z=z_c$ further reinforces our view that this value of z is in some deep sense critical. Varying p amounts to varying the stochasticity of the evolutionary process. That the value of $\sigma^2/N$ is independent of p at $z=z_c$ suggests that at $z=z_c$, there are critical fluctuations so that the value of p is largely irrelevant.

Given that these surprising and intricate dynamics occur in such a simple system, one is naturally led to ask whether they are general. If we believe that minority games capture some essential features of many complex adaptive systems, then we may seek to find counter-parts of the dynamics of these games in real social and biological systems. The period-two dynamics first identified in the non-evolutionary games[3] is a simple version of herding behavior which is very common in many social systems. Whether one can find manifestations of the more subtle evolutionary dynamics in real systems remains to be seen. But one should be optimistic, since it is almost certainly true that the simple structure of minority games play a role, however confounded by other effects, in a wide range of complex adaptive systems.

---

## Footnotes

[1] D. Challet and Y.-C. Zhang, *Physica A*, **246**, 407 (1997).

[2] R. Savit, R. Manuca andR. Riolo, Phys. Rev. Lett. **82**, 2203 (1999).

[3] R. Manuca, Y. Li, R. Riolo and R. Savit, *The Structure of Adaptive Competition in Minority Games*., UM Program for Study of Complex Systems Technical Report PSCS-98-11-001, at http://www.pscs.umich.edu/RESEARCH/pscs-tr.html, or LANL eprint at http://xxx.lanl.gov/abs/adap-org/9811005

[4] N. F. Johnson, M. Hart, and P. M. Hui, *Crowd effects and volatility in a competitive market*, LANL eprint at http://xxx.lanl.gov/abs/cond-mat/9811227

[5] M. A. R. de Cara, O. Pla, and F. Guinea, *Competition, efficiency and collective behavior in the "El Farol" bar model*, LANL eprint at http://xxx.lanl.gov/abs/cond-mat/9811162

[6] A. Cavagna, *Irrelevance of memory in the minority game*, LANL eprint at http://xxx.lanl.gov/abs/cond-mat/9812215.

[7] In ref. 3, we showed that the value of $z_c$ is closer to 0.35 than to 0.5, implying, for example, that $m_c$ is about 5.2 for N=101. Since the results presented in this paper were preformed for integer m, statements about specific value of $m_c$ should be understood to be approximate.

[8] Y.Li, R. Riolo and R. Savit, *Evolution in Minority Games II: Games with Variable Memory*, in preparation.

[9] Although not shown on the graph, $\sigma^2/N$ for evolutionary games with m=13 is about 0.23, and for m=14, is about 0.24. The RCG has a value of $\sigma^2/N$ of 0.25.

[10] With p=10% and m=3 and 4 convergence was slower. Those games were played for 600 generations.

[11] The asymptotic value of $\langle D_h \rangle$ depends on p, rising with increasing p. The example being discussed here is for p=20%.

[12] It is not difficult to see that for N=101, $D_b(i)$ will have a value of 50, for a distribution of random strategies, and will rise to 51 for an agent that is *always* in the minority group in a game in which the maximal number of points are distributed, so that the minority groups always have a population of 50 agents. In practice, absent evolutionary pressure, (i.e. during early generations) $D_b(i)$ more often falls in the range between 48 and 52, and late in evolution, when system-wide performance is fairly good, $D_b(i)$ is typically greater than about 50 and less than about 50.75.

[13] Notice also, that even if there were no selection against large values of $D_h(i)$, a depletion of agents with small values of $D_b(i)$ would also lead to a narrowing of the distribution of $D_b(i)$ from the high side: In order for some agents to have very high values of $D_b(i)$, there must also be some degree of clustering so that there are agents with very low values of $D_b(i)$. (To see this, consider the extreme example in which N-1 agents all have the same strategies, and one agent has the complement strategies.) As agents with low values of $D_b(i)$ are removed from the system so that clustering becomes less pronounced, it becomes increasingly unlikely to find agents with very high values of $D_b(i)$.

**Figure Captions**

Fig. 1 $\sigma^2/N$ vs z. for the adaptive, non-evolutionary case.

Fig. 2 $\sigma^2/N$ vs. m for with evolution N=101 and p=20%. There are eight independent runs for each value of m. Each point represents the value of $\sigma^2/N$ averaged over the last 50 generations of the run. Each generation is 10,000 time step, and runs were performed until $\sigma^2/N$ was sensibly constant, within fluctuations, generally about 200 generations. The horizontal dashed line is at the value of $\sigma^2/N$ for the RCG.

Fig. 3 $\sigma^2/N$ vs. z for different N and m and p=20%. Each point represents the value of $\sigma^2/N$ averaged over the last 50 generations of the run. Each generation is 10,000 time step, and runs were performed until $\sigma^2/N$ was sensibly constant, within fluctuations, generally about 200 generations. The horizontal dashed line is at the value of $\sigma^2/N$ for the RCG.

Fig. 4 $\sigma^2/N$ vs. m for N=101 and p=10%, 20% and 40%. Each point represents the value of $\sigma^2/N$ averaged over the last 50 generations of the run. Each generation is 10,000 time step, and runs were performed until $\sigma^2/N$ was sensibly constant, within fluctuations, generally about 200 generations.

Fig. 5 a. $\sigma^2/N$ as a function of generation for m=3, N=101 and p=10%, 20% and 40%.
b. $\sigma^2/N$ as a function of generation for m=7, N=101 and p=10%, 20% and 40%.

Fig. 6 a. $\sigma^2/N$ as a function of m, N=101, p=20%, s=1, $1^{st}$ generation. There are eight independent runs for each value of m.
b. $\sigma^2/N$ as a function of m, N=101, p=20%, s=1, $200^{th}$ generation. There are eight independent runs for each value of m.

Fig. 7 a. $\langle D_h \rangle$ as a function of generation for m=3 N=101, p=20%.
b. $\langle D_b \rangle$ as a function of generation for m=3 N=101, p=20%.
c. $\sigma^2/N$ as a function of generation for m=3 N=101, p=20%.

Fig. 8 a. $P(1|u_m)$ for the first generation for m=3 N=101, p=20%.

b. $P(1|u_m)$ for the $400^{th}$ generation for m=3 N=101, p=20%. Note that the histogram is very nearly, but not quite, flat.

Fig. 9 POED as a function of m for the $200^{th}$ generation in games played with N=101, s=2.

Fig. 10 a. Agent wealth vs. $D_h(i)$ every 20 generations for m=3, N=101, p=20%.

b. Agent wealth vs. $D_h(i)$ every $3^{rd}$ generation for the first 30 generations, for m=3, N=101, p=20%.

Fig. 11 a. Agent wealth vs. $D_b(i)$ every 20 generations for m=3, N=101, p=20%.

b. Agent wealth vs. $D_b(i)$ every $3^{rd}$ generation for the first 30 generations, for m=3, N=101, p=20%.

Fig. 12 $D_b(i)$ vs. $D_h(i)$ for m=3, N=101, p=20%, every 20 generations.

Fig. 13 The standard deviation of $D_b(i)$ as a function of generation for m=3, N=101, p=20%.

Fig. 14 a. 3-d plot of agent wealth vs. $D_h(i)$ and $D_b(i)$ after the first generation for m=3, N=101, p=20%.

b. 3-d plot of agent wealth vs. $D_h$ and $D_b$ after the last ($400^{th}$) generation for m=3, N=101, p=20%. Note the different $D_b$ and wealth scales in Figs. 14a and b.

Fig. 15 a. $\langle D_h \rangle$ as a function of generation for m=7 N=101, p=20%.

b. $\langle D_b \rangle$ as a function of generation for m=7 N=101, p=20%.

c. $\sigma^2/N$ as a function of generation for m=7 N=101, p=20%.

Fig. 16 Agent wealth vs. $D_h(i)$ every 20 generations for m=7, N=101, p=20%.

Fig. 17 Agent wealth vs. $D_b(i)$ every 20 generations for m=7, N=101, p=20%.